\documentclass[12pt,a4]{article}
\usepackage{amsthm,amsmath,latexsym,amssymb,amsfonts,amscd}
\usepackage{graphics,lscape,fancyhdr,array,stmaryrd,euscript}
\usepackage{ifthen,empheq,slashed}
\usepackage{sidecap}
\usepackage{ifpdf}
\usepackage{verbatim}
\usepackage{color}
\usepackage{relsize}
\usepackage[utf8]{inputenc}

\numberwithin{equation}{section}

\newcommand{\bep}{\begin{picture}}
\newcommand{\eep}{\end{picture}}

\newcommand{\smallpic}[1]{{\unitlength=0.2mm#1}}

\newcounter{YoungHeight}\newcounter{YoungWidth}

\newcounter{Mul1}\newcounter{Mul2}\newcounter{Mul3}\newcounter{Mul4}
\newcounter{A0}\newcounter{A1}\newcounter{A2}
\newcounter{B3}
\newcounter{C3}\newcounter{C4}
\newcounter{D1}\newcounter{D2}\newcounter{D3}
\newcounter{T0}\newcounter{T1}

\newlength{\txtHShift}

\newlength{\txtWidth}

\newcommand{\Add}[3]{\setcounter{#1}{#2}\addtocounter{#1}{#3}}

\newcommand{\Length}[1]{#10}
\newcommand{\YoungScale}{}

\newcommand{\BlockApar}[2]{\parbox{\Length{#1}pt}{\YoungScale\bep(\Length{#1},\Length{#2}){\Add{A1}{#1}{1}\Add{A2}{#2}{1}}%
\multiput(0,0)(10,0){\value{A1}}{\line(0,1){\Length{#2}}}\multiput(0,0)(0,10){\value{A2}}{\line(1,0){\Length{#1}}}%
\setcounter{YoungHeight}{\Length{#2}}\setcounter{YoungWidth}{\Length{#1}}\eep}}

\newcommand{\YoungpAA}{\BlockApar{1}{2}}

\newcommand{\YoungpAAAA}{\BlockApar{1}{4}}

\newcommand{\BlockAm}[2]{\mbox{\smallpic{\YoungScale\bep(\Length{#1},\Length{#2}){\Add{A1}{#1}{1}\Add{A2}{#2}{1}}%
\multiput(0,0)(10,0){\value{A1}}{\line(0,1){\Length{#2}}}\multiput(0,0)(0,10){\value{A2}}{\line(1,0){\Length{#1}}}%
\setcounter{YoungHeight}{\Length{#2}}\setcounter{YoungWidth}{\Length{#1}}\eep}}}

\newcommand{\YoungmB}{\BlockAm{2}{1}}

\newcommand{\YoungmAA}{\raisebox{-2pt}{\BlockAm{1}{2}}}

\usepackage{hyperref}

\ifpdf%
\usepackage[%
	sorting=none,%
	maxnames=199,%
	firstinits=true,%
	doi=false,%
	citestyle=numeric-comp,%
	backend=bibtex%
	]{biblatex}%
\AtEveryBibitem{\clearfield{title}}%
\renewbibmacro{in:}{}%
\else%
\fi%

\newcommand{\pl}{\partial}

\newcommand{\be}{\begin{equation}}
\newcommand{\ee}{\end{equation}}

\newcommand{\mm}{{\ensuremath{\underline{m}}}}
\newcommand{\nn}{{\ensuremath{\underline{n}}}}

\newcommand{\fud}[2]{{}^{#1}{}_{#2}\,}
\newcommand{\fdu}[2]{{}_{#1}{}^{#2}\,}

\newcommand{\besubeqs}{\begin{subequations}}
\newcommand{\esubeqs}{\end{subequations}}

\newcommand{\Fron}{{\Phi}}

\newcommand{\cons}[1]{\mathbb{D}_{#1}}
\newcommand{\tomd}{{\mathrm{d}}}

\newcommand{\Jsym}{J^{\YoungmB}}
\newcommand{\Jasym}{J^{\YoungmAA}}
\newcommand{\Jsing}{J^{\bullet}}
\newcommand{\JJst}[1]{{\overline{\langle J_{#1} J_{#1}\rangle}}}
\newcommand{\JOOst}[2]{{\overline{\langle J_{#1} O_{#2}O_{#2}\rangle}}}
\newcommand{\OOst}[1]{{\overline{\langle O_{#1} O_{#1}\rangle}}}

\begin{document}
\hfill
\begin{flushright}    
  {LMU-ASC 80/15}
\end{flushright}
\vskip 0.1\textheight

\begin{center}

{\Large\bfseries On (Un)Broken Higher-Spin Symmetry in Vector Models
} \\
\vspace{0.3cm}

\vskip 0.1\textheight

Evgeny \textsc{Skvortsov}${}^{1,2}$

\vspace{2cm}

{\em ${}^{1}$ Arnold Sommerfeld Center for Theoretical Physics\\
Ludwig-Maximilians University Munich\\
Theresienstr. 37, D-80333 Munich, Germany}\\
\vspace*{5pt}
{\em ${}^{2}$ Lebedev Institute of Physics, \\ Leninsky ave. 53, 119991 Moscow, Russia}

\vskip 0.05\textheight

{\bf Abstract }

\end{center}
\begin{quotation}
\noindent
The simplest consequences of exact and broken higher-spin symmetry are studied. The one-loop anomalous dimensions of higher-spin currents are determined from the multiplet recombination in the spirit of the modern bootstrap programme: the Wilson-Fisher CFT is studied both in the $4-\epsilon$-expansion and in the large-$N$. The bulk implications are briefly addressed: part of the higher-spin theory cubic action is reconstructed; one-loop corrections to the AdS masses of higher-spin  fields are discussed.
\end{quotation}

\newpage

\tableofcontents
\section{Introduction}

Wilson-Fisher CFT's describe many second-order phase transitions in the real world: vapor-water critical point, super-fluid $\lambda$-point, Ising model at the Curie temperature and many others. Wilson-Fisher CFT's remain unsolved, though they can consistently be approached by various expansion schemes: $4-\epsilon$ \cite{Wilson:1971dc} and large-$N$ \cite{Abe:1972,Vasiliev:1981yc,Lang:1990re}. Also, the numerical bootstrap programme \cite{ElShowk:2012ht} allowed to determine some of the anomalous dimensions to a very high precision \cite{El-Showk:2014dwa}.

With the dawn of the AdS/CFT correspondence \cite{Maldacena:1997re,Gubser:1998bc,Witten:1998qj} there had been attempts to find models that are simpler than full string theory vs. $\mathcal{N}=4$ SYM \cite{Sundborg:2000wp,Sezgin:2002rt}. According to the Klebanov-Polyakov conjecture \cite{Klebanov:2002ja}, see also \cite{Sezgin:2003pt,Leigh:2003gk}, the large-$N$ Wilson-Fisher CFT should be dual to the four-dimensional higher-spin theory \cite{Vasiliev:1990en}. With the different choice of the boundary conditions the same higher-spin (HS) theory should be dual to just a free scalar --- free $O(N)$ model.  First remarkable tests of this HS AdS/CFT duality have been performed in \cite{Giombi:2009wh} with a precursor in \cite{Sezgin:2003pt}. Recently, the HS AdS/CFT duality has successfully passed the one-loop tests in \cite{Giombi:2013yva,Tseytlin:2013jya,Giombi:2013fka, Giombi:2014yra,
Beccaria:2014jxa,Beccaria:2015vaa} and references therein/thereon.

The higher-spin theories are almost uniquely fixed by the infinite-dimensional gauge symmetries thereof \cite{Vasiliev:1988sa, Vasiliev:1989yr}, which come from gauging the higher-spin algebras \cite{Fradkin:1986ka}. Therefore, through the AdS/CFT duality the Wilson-Fisher CFT should inherit higher-spin symmetries as global ones, which can then be viewed as an infinite-dimensional extension of the conformal symmetry.

It turns out that the higher-spin (HS) symmetry works in a way that differs from its two-dimensional cousin Virasoro. When some HS algebra is a symmetry of a CFT, which means that there are conserved tensors $\pl^m J_{ma_2...a_s}=0$ on top of the stress-tensor, one can prove that the CFT is a free one \cite{Maldacena:2011jn,Alba:2013yda,Boulanger:2013zza,Stanev:2013qra,Alba:2015upa}.\footnote{There is a number of earlier works addressing the same issue \cite{Anselmi:1998ms} from a different perspective \cite{Fradkin:1986ka}.}
Moreover, all the correlation functions are given by the HS algebra invariants \cite{Colombo:2012jx, Didenko:2012tv}. This is the case for the duality with the free $O(N)$ model.

In Wilson-Fisher CFT the HS symmetry is broken or deformed, with the non-conservation of the higher-spin tensors having the specific form $\pl \cdot J=g JJ$ of double-trace operators, see e.g. \cite{Girardello:2002pp,Giombi:2011kc,Giombi:2012ms,Maldacena:2012sf}. In particular, in \cite{Maldacena:2012sf} it was shown that the broken higher-spin symmetry is powerful enough as to fix all the three-point functions up to few numbers to the leading order in $g$.

In the present paper we extract the anomalous dimensions from the non-conservation of the higher-spin currents to the first order in the coupling constant. To do so we need to work out the non-conservation operators $JJ$ and compute the correlation function thereof. There is a number of analytical studies appeared recently  \cite{Maldacena:2012sf,Komargodski:2012ek,Alday:2015ota,Rychkov:2015naa,Manashov:2015fha} that pursue new and closely related methods. In particular, \cite{Rychkov:2015naa} suggested to use the multiplet recombination, i.e. quantum equations of motion. In the case of HS currents the non-conservation operator $JJ$ recombines with the short multiplet of the HS current $J_{a_1...a_s}$ to form a long one, which is a non-conserved current. While phenomenologically HS currents may not be of much interest except for the few lower ones, it is the presence of an infinity of such 'almost conserved' HS tensors that is the signature of Wilson-Fisher CFT's and it is these currents that are dual to the HS fields of the HS theory in AdS. Therefore, it is important to zoom in on this subsector of the CFT.

The outline is as follows. In Section \ref{sec:Unbroken} we review the case of unbroken HS symmetries and reveal some of the simplest constraints that conserved higher-spin tensors impose on a CFT. Free boson and fermion are considered as examples and the relevant correlation functions of the higher-spin currents are computed. Next, in Section \ref{sec:HSbreaking} we review the two mechanisms of HS symmetry breaking: classical and quantum; derive the non-conservation operators for $4-\epsilon$ and large-$N$ Wilson-Fisher CFT and extract the anomalous dimensions of the HS currents. In the last Section \ref{sec:ads} we address several application for the AdS/CFT and the dual higher-spin theory: fix a part of the cubic action and make a conjecture for the full answer, suggesting that the duals of the free boson and fermion are the same in $AdS_4/CFT^3$; discuss one-loop corrections to the $AdS_{5-\epsilon}$-masses of higher-spin fields. In Appendices \ref{app:Thomas} and \ref{app:models} we review basic facts about CFT correlators and Wilson-Fisher CFT's. In Appendix \ref{app:Duals} we show how to extend the Vasiliev-type theories with the duals of higher-trace operators to the lowest order.

\section{Unbroken HS symmetry}\label{sec:Unbroken}
In the present section we consider unbroken HS symmetry whose signature is the existence in a CFT of at least one HS conserved current on top of the stress-tensor\footnote{The term HS current is a bit unfortunate in this context, because what we have is a HS conserved tensor, which can be used to construct HS currents by contracting it with Killing tensors. We will continue loosely refer to all of them as HS currents.}\footnote{The indices $a,b,...$ are those of the $d$-dimensional flat space where a CFT lives. A group of $s$ totally-symmetric indices $a_1...a_s$ is abbreviated as $a(s)$.}
\begin{align}
\pl^b J_{ba(s-1)}(x)=0\,, && s>2\,. \label{HScurrent}
\end{align}
First, we review the implications of unbroken HS symmetries --- the constraints imposed on a CFT by HS currents, simplest HS Ward identities. Then, we consider  free boson and free fermion as examples of CFT's with HS symmetries, which is aimed at collecting two- and three-point functions needed for the study of HS breaking and for the reconstruction of the AdS/CFT duals.

\subsection{Constraints by Unbroken HS symmetry}

Our main assumption is the existence of at least one HS current \eqref{HScurrent} in a CFT, where by CFT we mean all the usual conditions, of which the most important for us is the presence of the unique stress-tensor.

\paragraph{Simplest implications of HS symmetry.} For the case of the $2d$ minimal models the decoupling of a Virasoro singular vector imposed on the three-point function $\langle O_{\Delta_1}O_{\Delta_2}O_{\Delta_3}\rangle$ constrains $\Delta_i$. Likewise, the decoupling of the divergence of the HS current imposed on the simplest three-point functions results in important hints. First of all, HS currents can relate scalar operators of the same dimension only:\footnote{The structure of the two- and three-point functions in CFT and the useful technique of Thomas derivative to deal with tensorial primaries are reviewed in Appendix \ref{app:Thomas}. In practice the conservation is imposed with the help of the Thomas derivative and we denote the operator that checks the conservation of the $i$-th operator as $\cons{i}$ in the index free notation, which in components reduces to  \eqref{HScurrent}. }
\begin{align}
\cons{1}\langle J_s O_{\Delta_1} O_{\Delta_2}\rangle&=0 && \Longrightarrow && \Delta_1=\Delta_2\,. \label{equaldims}
\end{align}
This fact is already true for the global symmetry current $s=1$ and the stress-tensor $s=2$, see e.g. \cite{Fradkin:1996is}. Secondly, the three-point function of two HS currents and one scalar operator leads to\footnote{Here we assumed parity and we are in generic dimension. If all possible conformal structures are taken into account it is possible to see free fermion, etc. See \cite{Giombi:2011rz} for some $3d$ examples. }
\begin{align}
\cons{1,2}\langle J_s J_{s'}O_{\Delta} \rangle&=0 && \Longrightarrow && \Delta=d-2\,.
\end{align}
This dramatic restriction requires a genuine HS current and is not true for the $s=1,2$ currents that can have scalars of arbitrary dimension in the OPE. The value of $d-2$ suggests that $O$ is nothing but $\phi^2$ for a free scalar $\phi$ and the presence of a free field is not what one would expect from a  nontrivial CFT.

\paragraph{From OPE to HS algebra.} The OPE is the basic tool to study CFT's. For operators carrying nontrivial tensor representations, e.g. HS currents, the OPE can be quite complicated. In general in the OPE of two HS currents  we expect to find
\begin{align}
J_{s_1}J_{s_2}=\delta_{s_1,s_2} \langle J_{s_1}J_{s_2}\rangle + \sum J_k +O_2\,,
\end{align}
where there is a two-point function, a sum over HS currents of different spins and, possibly, the spin-zero $J_0$ operator may also be present. Some other operators, collectively denoted as $O_2$ must also be present as the $N$-counting shows \cite{Klebanov:2002ja}. For example, the normal product $:J_{s_1}J_{s_2}:$ should be there. For the purpose of studying the HS currents content the structure of the OPE can be simplified by integrating over the insertion of $J_{s_1}$ to form a HS charge. First, one builds a set of currents by contracting a HS conserved tensor with a conformal Killing tensor $v^{a(s-1)}$:
\begin{align}
j_m(v)&=J_{ma(s-1)}v^{a(s-1)}\,, && \pl^a v^{a(s-1)}-\text{traces}=0\,,
\end{align}
which applies to the stress-tensor as well, for which case Killing tensor becomes Killing vector.
Such currents are conserved in the usual sense of $\pl^m j_m(v)=0$ and can be integrated to form a charge
\begin{align}
Q(v)&= \int dS^m j_{m}(v)\,.
\end{align}
When applied to the OPE one gets the action of the HS charge on the HS current:
\begin{align}
[Q_{s_1}, J_{s_2}]&= \sum_s J_{s} \label{HSreptheory}\,,
\end{align}
which is nothing but the action of the HS algebra formed by $Q_{s}$ for all spins $s$ present in the spectrum on the HS currents themselves. This was the main starting point in \cite{Maldacena:2011jn}. One can make one step further in abstracting the HS algebra --- to integrate the second time, which leads to the Lie bracket of two HS charges:
\begin{align}
[Q_{s_1},Q_{s_2}]&=\sum Q_k \label{HSalgebra}\,.
\end{align}
The commutator has to obey the Jacobi identity and the algebra formed by $Q_s$ contains the conformal algebra realized by the stress-tensor via $Q_2$ and at least one genuine HS charge $Q_s$. Then the Jacobi identity can be solved. This was studied in \cite{Boulanger:2013zza}. The Jacobi identity is purely algebraic and does not appeal to any local field realization of the HS currents.

The main result of studying HS algebras via the Lie bracket \eqref{HSalgebra} or the representation theory via \eqref{HSreptheory} is that \cite{Maldacena:2011jn,Alba:2013yda,Boulanger:2013zza,Stanev:2013qra,Alba:2015upa}: (i) there are infinitely many of HS currents/charges once at least one is present; (ii) the correlators are those of one of the free CFT's.

Let us note that the unitarity does not matter much for our considerations. In particular, the presence of at least one HS current still requires all of them to be present. As an example, one can consider free fields obeying higher order equations $\square^k\phi=0$, $k>1$, \cite{Bekaert:2013zya}, the HS currents are still there. In the free theory the OPE is just Wick's theorem. The full OPE structure for some of the free CFT's was explicitly worked out in \cite{Gelfond:2013xt}, where one can see how the action of the HS charges leads to HS algebras via Wick's theorem.

\paragraph{Simplest HS Ward Identities.} Many of the results known for the case of $s=1$ global symmetry currents and $s=2$ stress-tensor can be generalized to HS currents. As was discussed above the OPE contains the complete information about HS symmetries. Let us study the simplest implications of the HS Ward identities. The basic Ward identity in the integral form for a symmetry $\delta^v$ generated by current $j_m(v)$ is
\begin{align}\label{WardIdentity}
-\int dS^m\, \langle j_m(v) O(x_1)... \rangle=\sum_i \langle... \delta^v_i O(x_i)...\rangle\,.
\end{align}
A useful example is provided by the free scalar field $\phi(x)$ in $d$-dimensions. There is a lot of HS transformations generated by various Killing tensors contracted with the same HS current. One can consider the simplest instance given by hyper-translations $\delta_v =v^{a(s-1)}\pl_{a(s-1)}$ for $v^{a(s-1)}$ constant. The action of the spin-$s$ charge on $\phi$ and $\phi^2$ are
\begin{align}
[Q_s, \phi]&=v^{a(s-1)}\pl_{a(s-1)}\phi\\
[Q_s, \phi^2]&= \pl^{s-1}\phi^2 +\pl^{s-3}J_2+...+J_{s-1}\label{HStransA}
\end{align}
where in the last line we just sketched the general structure, see e.g. \cite{Maldacena:2011jn}. The first formula is self-evident since $\delta \phi= v^{a(s-1)}\pl_{a(s-1)}\phi$ is a symmetry of the Klein-Gordon equation $\square \phi=0$ for $v^{a(s-1)}$ constant. It is also the symmetry of the action $\tfrac12\int (\pl\phi)^2$.

Contrary to the spin-two and spin-one charges the action of the HS charges may involve an infinite number of fields. Indeed, the global symmetry or the conformal symmetry slices the totality of fields into irreducible representations of a given symmetry. The HS algebra is infinite-dimensional. Its smallest representations are the identity and the free field $\phi$ itself, which plays the role of the fundamental one. Other representations are given by tensor powers of $\phi$ and decompose into infinite number of conformal fields. The simplest case is the tensor square $\phi\otimes\phi$ and the Flato-Fronsdal theorem \cite{Flato:1978qz} tells that it decomposes into all HS currents $J_0\oplus J_2 \oplus J_4\oplus...$ with $J_0=\phi^2$ being the first degenerate member.\footnote{The Flato-Fronsdal theorem was generalized to any $d$ in \cite{Vasiliev:2004cm}.}

When applied to three-point functions $\langle J_s OO\rangle$ the Ward identity \eqref{WardIdentity} relates them to  two-point functions $\langle OO\rangle$, which allows to fix the coupling constants. Indeed, the three-point function of a HS current and a weight-$\Delta$ scalar operator $O_\Delta$ is fixed by conformal symmetry up to a number $g_{s00}$, the coupling constant:
\begin{align}
\langle J_s O_\Delta O_\Delta\rangle&= g_{s00}\JOOst{s}{\Delta}\,,
\end{align}
where we introduce the standard conformal structure:
\begin{align}\label{standardthreepoint}
\JOOst{s}{\Delta}&=(Q\cdot \xi)^s \left(\frac{x_{23}^2}{x_{12}^2x_{13}^2}\right)^{\frac{d-2}2}\frac{1}{(x_{23}^2)^\Delta}\,.
\end{align}
Here, $\xi^a$ is light-like auxiliary vector that projects out traces, see Appendix \ref{app:Thomas}, and the correlator is factorized via the conformally-covariant vector
\begin{align}
Q^a&=\left(\frac{x^a_{21}}{x_{21}^2}-\frac{x^a_{31}}{x_{31}^2}\right)\,.
\end{align}
Likewise, the two-point function $\langle O_\Delta O_\Delta\rangle$ is fixed up to a number $C_{OO}$ to be the standard conformal structure:
\begin{align}\label{stadardtwopoint}
\langle O_\Delta(x_1) O_\Delta(x_2)\rangle&=C_{OO}\OOst{\Delta}\,,&& \OOst{\Delta}=\frac{1}{(x^2_{12})^\Delta}\,.
\end{align}
Couplings $g_{s00}$ are related in a simple way to the OPE coefficients of $J_s$ in the $OO$ OPE. In order to fix the coupling constants $g_{s00}$ we can consider the simplest Ward identity for the hyper-translations. By integrating over a small ball around the insertion of the first operator, which is conveniently placed at $y=0$, we should get:\footnote{The integral form of the Ward identity is easier to work with since the integrals are well-defined. In the differential form of the Ward identities one has to regularize the otherwise ill-defined distributions. For example, one can allow one of the fields to have an infinitesimal anomalous dimension or directly apply differential regularization. }
\begin{align}
\xi^{a}...\xi^a\int \pl^m{\langle J_{ma(s-1)} O_\Delta(0)O_\Delta(z)\rangle}=\xi^{a}...\xi^a\pl^y_{a}...\pl^y_a\langle O_\Delta(y)O_\Delta(z)\rangle\big|_{y=0}\,.
\end{align}
The right-hand side is simple while the integral can be done by replacing the correlation function by its actual value with a unit coefficient in front of the standard structure:
\begin{align}
\xi^{a}...\xi^a\int \pl^m\overline{\langle J_{ma(s-1)} O_\Delta(0)O_\Delta(z)\rangle}&=S_d f_s(\xi\cdot z)^{s-1}\frac{1}{(z^2)^{\Delta+s-1}}\,,\\
\xi^{a}...\xi^a \pl^y_{a}...\pl^y_a\langle O_\Delta(y)O_\Delta(z)\rangle\big|_{y=0}&=\frac{ 2^{s-1}\Gamma[\Delta+s-1]}{\Gamma[\Delta]}(\xi\cdot z)^{s-1}\frac{1}{(z^2)^{\Delta+s-1}}\,,
\\
f_s&=\frac{ 2^{d+2 s-4} \Gamma \left(\frac{d}{2}\right) \Gamma \left(\frac{d-3}{2}+s\right)}{\sqrt{\pi } (d+2 s-4) \Gamma (s) \Gamma (d+s-3)}\,,
\end{align}
where $S_d$ is the area of the $d$-dimensional sphere, $S_d=2\pi^{\tfrac{d}2}/\Gamma[\tfrac{d}2]$. Now we can express all the cubic couplings $g_{s00}$ in terms of the two-point normalization factor $C_{OO}$:
\begin{align}
g_{s00}&=\frac{C_{OO}}{S_d F_s}\,, && F_s=\frac{f_s\Gamma[\Delta]}{2^{s-1}\Gamma[\Delta+s-1]}\,.
\end{align}
In the case of $s=1,2$ the above formula boils down to the well-known, see e.g. \cite{Osborn:1993cr}, relations
\begin{align}
g_{100}&=\frac{C_{OO}}{S_d}\,, &
g_{200}&=\frac{C_{OO}d\Delta}{(d-1)S_d}\,.
\end{align}
For example, for the free $3d$ scalar and fermion --- weight-one $\phi^2$ and weight-two $\bar{\psi}\psi$ scalar operators coupled to HS currents of even spins --- we find
\begin{align}
\Delta=1&: && g_{s00}=C_{OO}2^{1-s} (2 s-1) \Gamma (s)^2\,,\\
\Delta=2&: && g_{s00}=C_{OO} 2^{1-s} (2 s-1) \Gamma (s)^2 s\,,
\end{align}
the ratio being $1/s$. Let us note that the Ward identity for the HS symmetry we studied above is the simplest one. There are Ward identities with more than one HS currents, which allows to uncover the full HS algebra, as was done in \cite{Maldacena:2011jn} for the $3d$ case.

\paragraph{HS Invariants.} Since many of the coefficients depend on the way the two- and three-point functions are normalized it is important to pass to invariants. For example, the normalization of the two-point functions is arbitrary unless it is a global symmetry current or the stress-tensor, whose normalization is fixed by the Ward identities and contain important information, which is called central charges.  There is only one invariant that is relevant for $\langle J_s J_0 J_0\rangle$:
\begin{align}
I_{s00}&=\frac{\langle J_s J_0 J_0\rangle^2}{{\langle J_s J_s\rangle} \langle J_0 J_0\rangle^2}\label{HSinvariant}\,,
\end{align}
where it is assumed that some standard representatives of the correlation functions are chosen, e.g. \eqref{stadardtwopoint}, \eqref{standardthreepoint} and we take the ratio of the factors multiplying those. In terms of the HS algebra structure constants $f_{s_1,s_2,s_3}$ and the invariant metric $g_{s_1,s_2}\sim \delta_{s_1,s_2}$ defined by the three- and two-point functions, respectively, the ratio above corresponds to $f^2_{s00}{g^{ss}} g^{00}g^{00}$. In case of the free $O(N)$ boson/fermion this invariant scales as $N^{-1}$ with $N$, as will be shown below.

\paragraph{HS central charges.} In a given CFT one might have to rescale the operators as to have the Ward identities satisfied with the canonical normalization. This can be done by replacing $J_s$ with $J_s {\lambda_s}$ where $\lambda_s=C_{OO}/(F_s S_d g_{s00})$. As a result the Ward identities look canonical, but the two-point function of the HS currents gets rescaled and is expressed in terms of the invariant:
\begin{align}
C_{ss}&= \frac{1}{I_{s00}F_s^2 S_d^2}\,.
\end{align}
The HS central charges \cite{Anselmi:1998bh} can be defined by analogy with the global symmetry and the stress-tensor central charges as:
\begin{align}
\langle J_s J_s \rangle&= \frac{C_{J,s}}{S_d^2} \frac{(P_{12})^s}{(x_{12}^2)^{d+s-2}}\,, && C_{J,s}=\frac{1}{I^2_{s00}F_s^2}\,,
\end{align}
where $P_{12}$ is the unique two-point conformally-covariant structure
\begin{align}
P_{12}&=\xi^1_a \xi^2_b\left(\delta^{ab}-2\frac{x_{12}^ax_{12}^b}{x_{12}^2}\right)\,.
\end{align}
$C_{J,s}$ scales as $N$ in the free vector models. Since the HS symmetries are present in free CFT's only, the HS central charges may play a less fundamental role and we use them to encode the same information as the invariants \eqref{HSinvariant} provided some standard normalization is chosen.

\subsection{Free Boson}
In $O(N)$ vector models the fundamental field is an $O(N)$ vector, the spin field $\phi^i(x)$, $i=1,...,N$. If the model is the free one or  the interacting one taken at the strict $N\rightarrow \infty$ limit, there is a set of conserved HS currents. The expressions for the HS currents have been obtained a countable number of times in the literature, see e.g. \cite{Craigie:1983fb}, but let us construct them once again by reiterating the method of \cite{Craigie:1983fb}. The most general distribution of derivatives over the two fields can be obtained via
\begin{align}\label{HScurrentscalar}
J^{ij}(x,\xi)&=F(\xi\cdot\pl_1,\xi\cdot \pl_2)\, :\phi^i(x_1)\phi^j (x_2): \Big|_{x_1=x_2=x}\,,
\end{align}
where we again prefer to hide all indices away by contracting them with an auxiliary vector $\xi^a$. It can be rewritten by stripping off the center of mass coordinates as
\begin{align}
F(u,v)&=  \sum_s (u+v)^s F_s\left(\frac{u-v}{u+v}\right)\,.
\end{align}
The HS currents are primaries and hence are traceless, which is taken into account by the light-like polarization vector $\xi^a$, $\xi\cdot\xi=0$. As a consequence, the conservation has to be imposed with the help of the Thomas operator, see Appendix \ref{app:Thomas}. The solution to the conservation constraint is rather simple ($\nu=(d-3)/2$):
\begin{align}
F(t,w)&=(1-2wt+t^2)^{-\nu}\,,  &F(t,w)&=\sum_s F_s(w)t^s\,,
\end{align}
and, as is well-known, yields Gegenbauer polynomials, $F_s(w)=C^\nu_s(w)$. Therefore, the generating function of all the HS currents is
\begin{align}
J^{ij}(x,\xi)&=\Big(1-2(\pl_1- \pl_2)(\pl_1+ \pl_2)+(\pl_1+ \pl_2)^2\Big)^{-\nu} \, :\phi^i(x_1)\phi^j (x_2): \Big|_{x_1=x_2=x}\,,
\end{align}
where we abbreviated $\pl_i=\xi\cdot \pl_i$.

In general HS currents carry a rank-two reducible representation of $O(N)$ and can be decomposed into three irreducible components:
\begin{align}
J^{ij}=\Jsym+\Jasym+\frac1{N}\Jsing\,,
&&
\begin{aligned}
\Jsym&= \frac12(J^{ij}+J^{ji})-\frac1{N} \delta^{ij}\delta_{kn}J^{kn}\,,\\
\Jasym&= \frac12(J^{ij}-J^{ji})\,,\\
\Jsing&=\delta_{kn}J^{kn}\,.
\end{aligned}
\end{align}
The behavior of the three components at the quantum level is quite different. The singlet sector, which the stress-tensor belongs to, is important for AdS/CFT.

In what follows we will need two- and three-point functions, which can be computed by simple Wick contractions of the spin field $\phi^i$. The canonically normalized two-point function of $\phi^i$ is
\begin{align}\label{phitwopoint}
\langle \phi^i(x_1) \phi^j(x_2)\rangle&=\delta^{ij}\frac{\Gamma[d/2-1]}{4\pi^{d/2} (x_{12}^2)^{d/2-1}}=\frac{\delta^{ij}}{4\pi^{d/2}} \int d\alpha\, \alpha^{d/2-2} e^{-\alpha x_{12}^2}\,,
\end{align}
where the latter expression is the most useful one as it allows to replace complicated tensor structures resulting from derivatives of the spin field with polynomials in the Schwinger parameters.

The two-point function $\langle J_s J_s\rangle$ can be straightforwardly evaluated with the orthogonality relation $\langle J_{s_1} J_{s_2}\rangle\sim \delta_{s_1,s_2}$ boiling down to that of Gegenbauer polynomials:
\begin{align}
\langle J^{ij}_s J^{kl}_s\rangle&= [\delta^{ik}\delta^{jl}+(-)^s\delta^{il}\delta^{jk}]C_s\times \JJst{s}\,,  \\
C_j&=\frac{\pi  2^{-2 d+j+8} \Gamma (d+j-3) \Gamma (d+2 j-2)}{\left(4 \pi ^{d/2}\right)^2 j! (d+2 j-3) \Gamma \left(\frac{d}{2}-\frac{3}{2}\right)^2}\,,
\end{align}
where we singled out the two-point structure with a unit coefficient
\begin{align}
 \JJst{s} &=\frac{(P_{12})^s}{(x_{12}^2)^{d+s-2}}\,, 
\end{align} which plays the role of the standard one. The two-point function can be projected onto the three irreducible $O(N)$-structures as:
\begin{align}
\langle \Jsing_s \Jsing_s\rangle&= N[1+(-)^s]C_s \JJst{s}\,,\label{scalarst} \\
\langle \Jasym_s \Jasym_s\rangle&= \frac12(\delta^{ik}\delta^{jl}-\delta^{il}\delta^{jk})[1-(-)^s]C_s \JJst{s}\,, \\
\langle \Jsym_s \Jsym_s\rangle&= \frac12 (\delta^{ik}\delta^{jl}+\delta^{il}\delta^{jk}-\frac{2}{N}\delta^{ij}\delta^{kl})C_s [1+(-)^s]\JJst{s}\,.
\end{align}
The two-point function of singlets scales as $N$ and the anti-symmetric currents are nontrivial for odd spins while the singlet and symmetric currents are so for even spins. Analogously, the three-point functions can be found to be normalized as
\begin{align}
\langle J^{ij}_s J_0 J_0\rangle&= 4\delta^{ij}[1+(-)^s] C_{sOO}\times \JOOst{s}{d-2}\,, \\
C_{sOO}&=\pi ^{-\frac{3 d}{2}} 2^{s-6} \Gamma \left(\frac{d}{2}-1\right)^2 \Gamma \left(\frac{d}{2}+s-1\right)\,,
\end{align}
where $4$ is because of the Wick contractions and we split off the standard conformal structure \eqref{standardthreepoint}. The three-point function of the spin field with the HS current is
\begin{align}
\langle J^{ij}_s \phi^k \phi^l\rangle&= [\delta^{ik}\delta^{jl}+(-)^s\delta^{il}\delta^{jk}] C_{s\phi\phi}\times \JOOst{s}{\tfrac{d-2}2}\,, \\
C_{s\phi\phi}&=\frac{\pi ^{-d} 2^{s-4} \Gamma \left(\frac{d}{2}-1\right) \Gamma \left(\frac{d}{2}+s-1\right) \Gamma (d+s-3)}{s! \Gamma (d-3)}\,.
\end{align}
The invariant \eqref{HSinvariant} built of the single-trace correlation functions \eqref{scalarst} is
\begin{align}
I_{s00}&=\frac{\sqrt{\pi } 2^{-d-s+7} \Gamma \left(\frac{d}{2}+s-1\right) \Gamma (d+s-3)}{N \Gamma \left(\frac{d}{2}-1\right)^2 \Gamma (s+1) \Gamma \left(\frac{d-3}{2}+s\right)}
\end{align}
and can also be extracted from \cite{Diaz:2006nm}. It is $I_{s00}$ that has to be reproduced by the dual HS theory upon plugging boundary-to-bulk propagators into the cubic vertices as will be discussed in Section \ref{sec:ads}. Combining the invariant with the discussion above we see that the HS central charge is
\begin{align}
C_{J,s}&=\frac{\sqrt{\pi } N 2^{-d-s+3} \Gamma (s)^2 \Gamma (s+1) \Gamma \left(\frac{d}{2}+s-1\right) \Gamma (d+s-3)}{\Gamma \left(\frac{d}{2}\right)^2 \Gamma \left(\frac{d-3}{2}+s\right)}\,,
\end{align}
where the formula works by construction for even spins. This formula agrees with the $s=2$ case, for which it gives the well-known \cite{Osborn:1993cr}
\begin{align}
C_T&=\frac{d N}{d-1}\,.
\end{align}

\subsection{Free Fermion}
As in any free CFT there are HS currents in the free fermion model too. The same idea as in the scalar case leads to
\begin{align}
J\fud{i}{j}(x,\xi)&=F(\xi\cdot\pl_1,\xi\cdot \pl_2)\, :\bar{\psi}^i(x_1)(\gamma \cdot \xi) \psi_j (x_2): \Big|_{x_1=x_2=x}\,,
\end{align}
where $i,j=1,..\tilde{N}$ are the $U(\tilde{N})$ indices and
\begin{align}
F(u,v)&=  \sum_s (u+v)^{s-1} F_s\left(\frac{u-v}{u+v}\right)\,.
\end{align}
The solution is again given by Gegenbauer polynomials with a bit different parameters ($\nu=\tfrac{d-3}{2}$):\footnote{In \cite{Anselmi:1999bb} another presentation for the HS currents was found, which up to an overall coefficient can be obtained by Taylor expanding the Gegenbauer polynomial.}
\begin{align}
F(t,w)&=t(1-2wt+t^2)^{-\nu-1}\,,  &F(t,w)&=\sum_s F_s(w)t^s\,.
\end{align}
In other words, $F_s(w)=  C^{\nu+1}_{s-1}(w)$ and
\begin{align}\notag
J\fud{i}{j}(x,\xi)&=(\pl_1+ \pl_2)\Big(1-2(\pl_1- \pl_2)(\pl_1+ \pl_2)+(\pl_1+ \pl_2)^2\Big)^{-\nu-1} \, :\bar{\psi}^i(x_1)(\gamma\cdot\xi) \psi_j (x_2): \Big|_{x_1=x_2=x}\,.
\end{align}
The two-point function of $\psi^i$ differs from that of the scalar field by an extra $\slashed{\pl}=\gamma\cdot \pl$:
\begin{align}
\langle {\psi}_j(x_1) \bar{\psi}^i(x_2)\rangle&=-\delta^i_j\frac{1}{S_d} \frac{\slashed{x}}{(x_{12}^2)^{d/2}}=\delta^i_j\slashed{\pl} \frac{\Gamma[\tfrac{d}{2}-1]}{4\pi^{d/2} (x_{12}^2)^{d/2-1}}=\slashed{\pl}\frac{\delta^{i}_{j}}{4\pi^{d/2}} \int d\alpha\, \alpha^{d/2-2} e^{-\alpha x_{12}^2}\,.
\end{align}
A new interesting feature is that the HS currents can have more complicated symmetry types, \cite{Vasiliev:2004cm,Alkalaev:2012rg,Alkalaev:2012ic}. Namely, we can replace $(\gamma \cdot \xi)$ with $\gamma^{u[q-1]v}\xi_v$ where $\gamma^{u[q]}$ is the anti-symmetrized product of $\gamma$-matrices and an additional projector is needed to make the currents Young-irreducible, i.e. quasi-primary. We will not consider such currents, restricting to the totally-symmetric ones, which are present in any free CFT.

The two-point function of the HS currents can be computed in full analogy to the free boson case. The $U(\tilde{N})$-decomposition can easily be worked out and we record for the future just the normalization of the singlet HS currents:
\begin{align}
\langle \Jsing_s \Jsing_s\rangle&= N C_s \JJst{s}\,, \\
C_j&=\frac{(-)^j\pi^{1-d}  2^{-2 d+j+1} \Gamma (d+j-2) \Gamma (d+2 j-3)}{\Gamma \left(\frac{d-1}{2}\right)^2 \Gamma (j)}\,.
\end{align}
where $N=\tilde{N} \text{tr} \boldsymbol{1}$ is the total number of fermions. The non-singlet currents have an obvious tensor structure and $\text{tr} \boldsymbol{1}$ instead of $N$. Note that $(-)^j$ is exactly what one needs for the reflection positivity, but it would be better to introduce $i^s$ into the currents. A word of warning is that the lowest singlet $O=J_0=\bar{\psi}\psi$ cannot be obtained from $J_s$ at $s=0$, but since the three-point function $\langle OOO\rangle=0$ we can ignore this case. After rescaling the generating function appropriately, this result matches the one in \cite{Anselmi:1999bb}, which was obtained by difficult resummation. Also, the three-point function is of interest
\begin{align}
\langle J_s J_0 J_0\rangle&= 2N C_{sOO}\times \JOOst{s}{d-1}\,, \\
C_{jOO}&=\frac{ \pi ^{-\frac{3 d}{2}} 2^{j-4} \Gamma \left(\frac{d}{2}\right)^2 \Gamma \left(\frac{d}{2}+j-1\right) \Gamma (d+j-2)}{\Gamma (d-1) \Gamma (j)}\,,
\end{align}
since it determines the coupling constants. It vanishes for odd spins, which is implicit in the formula. In $AdS_4/CFT_3$ duality the free boson and fermion have the same dual, in particular the HS algebras are the same \cite{Konstein:1989ij, Boulanger:2013zza}. The invariant $I_{s00}$ is
\begin{align}
I_{s00}=&\frac{\sqrt{\pi } (-)^{s} 2^{-d-s+5} \Gamma \left(\frac{d}{2}+s-1\right) \Gamma (d+s-2)}{N \Gamma \left(\frac{d}{2}\right)^2 \Gamma (s) \Gamma \left(\frac{d-3}{2}+s\right)}
\end{align}
and differs by $s(d+s-3)/(d-2)^2$ from that of the free boson. We will see in Section \ref{sec:ads} how this difference is compensated by changing the boundary conditions from free boson $\Delta=1$ to free fermion $\Delta=2$ for the $AdS$ duals of $O=\phi^2, \bar{\psi}\psi$. 

\subsection{HS singletons}
Lastly, we note that the generalization of the above formulae to the conformal fields with higher spin is obvious, see also \cite{Craigie:1983fb}. For example, in the case of the $n=(d-2)/2$-forms $A_{a[n]}$ in even dimension, of which  $4d$ Maxwell field $A_m$ is a particular case, we find that the totally-symmetric HS currents
\begin{align}
J&=F(\xi\cdot\pl_1,\xi\cdot \pl_2)\, \xi^a \xi^a:G_{ab[n]}(x_1)G\fdu{a}{b[n]} (x_2): \Big|_{x_1=x_2=x}
\end{align}
are generated by $C^{\nu+2}_{j-2}(w)$. Here $G_{a[n+1]}=\pl_{a}A_{a[n]}+\text{permutations}$ is the totally anti-symmetric field-strength. Similarly, the HS singletons \cite{Bekaert:2009fg} can be taken into account, the currents being constructed by $F=C^{\nu+2s}_{j-2s}(w)$
\begin{align}
J&=F(\xi\cdot\pl_1,\xi\cdot \pl_2)\, \xi^{a(2s)} :G_{a(s),b[n](s)}(x_1)G\fdu{a(s)}{b[n](s)} (x_2): \Big|_{x_1=x_2=x}\,,
\end{align}
where $G_{a(s)[n+1]}$ is the field-strength that has the symmetry of the rectangular $s\times n+1$ Young diagram. As is known, see \cite{Gelfond:2006be} for the explicit construction in $4d$, the lowest spin for the conserved tensor built out of a spin-$s$ conformal field is $2s$, which is also clear from the index of the Gegenbauer polynomial here-above.

\section{HS symmetry breaking}\label{sec:HSbreaking}
In this section we discuss two possible ways to break HS symmetries: classical and quantum, which are realized in the theories of Yang-Mills and Wilson-Fisher type.\footnote{This is how we distinguish between classical and quantum. It will be clear that the definitions are well-defined in a more broad sense when a CFT is weakly-coupled.} Then, we scrutinize the quantum breaking and study rather general implications of the conformal symmetry and quantum equations of motion for HS currents. The non-conservation operators for Wilson-Fisher CFT in $4-\epsilon$ and large-$N$ are derived and used to extract the anomalous dimensions of the HS currents at the first non-trivial order in the coupling constant. In Appendix \ref{app:models} we collected basic facts about the models studied in the main part.

\subsection{Classical and Quantum Breaking}
There are at least two mechanisms to break HS symmetries known. Due to AdS/CFT each way of breaking has both CFT and AdS interpretation. When the HS symmetry is exact the conservation of a HS current in a free $CFT_d$ is dual to the gauge transformations for the Fronsdal field \cite{Fronsdal:1978rb} in $AdS_{d+1}$:\footnote{The indices $\mm,\nn,...=0,...,d$ are the indices of $AdS_{d+1}$ tensors.}
\begin{align}
\pl^m J_{ma(s-1)}&=0 &&  \Longleftrightarrow &&\delta\Fron_{\mm(s)}=\nabla_\mm \xi_{\mm(s-1)}+O(z,\Phi\xi)\,.
\end{align}
This is true near the boundary. When the bulk field departs from the $AdS$ boundary at $z=0$ interactions switch on and the free gauge symmetries get deformed by the terms of the form $\Phi\xi$ (like in Yang-Mills and gravity) and more complicated and nonlinear ones (as different from Yang-Mills and gravity). The duals of the free CFT's are HS theories as was discussed in \cite{Sundborg:2000wp,Klebanov:2002ja,Sezgin:2002rt,Didenko:2012vh}. HS theories should gauge the HS algebra of a given free CFT, but at present only the dual of the free boson is known in any $AdS_{d+1}$ \cite{Vasiliev:2003ev} and the dual of the free fermion is available in $AdS_4$ thanks to the already mentioned coincidence that the HS algebra of the free boson is identical to that of the free fermion in $3d$. The same HS theories can be the duals of interacting CFT's by changing the boundary conditions within the unitarity window \cite{Klebanov:2002ja,Sezgin:2003pt,Leigh:2003gk}.

When HS symmetry is broken the conservation of the HS currents is replaced by non-conservation, the structure being dependent on the mechanism of breaking. Accordingly, the dual bulk fields have to acquire a mass, which can be done classically \cite{Bianchi:2005ze} and then followed by radiative corrections or immediately via loops \cite{Girardello:2002pp}.

\paragraph{Classical, BEH.} One can think of a HS generalization of the usual Brout-Engler-Higgs mechanism --- a massless vector boson swallows the Goldstone scalar which makes the right number of degrees of freedom to become a massive vector. In the case of the original particle being a massless spin $s>1$ field in $AdS_{d+1}$ the correct partner is a massive spin-$(s-1)$ field of mass $M^2=(d+s-2)(s-1)$ in the units of the cosmological constant.

The field realization of the HS BEH mechanism is nicely incorporated into the Lagrangian description of massive HS fields proposed by Zinoviev \cite{Zinoviev:2001dt}. The idea is to take the sum of the Fronsdal Lagrangians for $j=0,1,...,s$, which has the right total number of degrees of freedom to describe a single massive spin-$s$ field. All possible mixings and mass-like terms are then added to make the Lagrangian irreducible, as well as the gauge transformations $\delta \Fron_j=\nabla \xi_j$ are appended with the gauge parameters of fields with neighboring spins:
\begin{align}
\mathcal{L}_{s,m^2}&=\sum_{j=0}^{j=s}\mathcal{L}_j +\sum_j\left[\Fron_j \nabla \Fron_{j-1} + m_j^2 \Fron_j^2\right]\,, && \delta\Fron_j=\nabla \xi_j+ \xi_{j+1}+ \eta\xi_{j-1}\,,
\end{align}
where $\eta$ is the $AdS$ metric. The gauge invariance fixes all the free coefficients in terms of a single constant, which can be identified with the mass. At the special point where the $AdS$-mass is that of the spin-$s$ massless field the Lagrangian splits into two parts: the Fronsdal one for a spin-$s$ massless field and the Lagrangian for a massive spin-$(s-1)$ field with mass $M^2$:
\begin{align}
\mathcal{L}_{s,m^2}\Big|_{m^2=(d+s-2)(s-2)-s}&=\mathcal{L}_s+\mathcal{L}_{s-1,M^2}\,.
\end{align}

On the CFT side, the HS current non-conservation is determined by some single-trace operator present in the theory:
\begin{align}
\cons{}J&=g O_1\,.
\end{align}
A good example is Yang-Mills theory. At the zero coupling it has a HS symmetry manifested by $J_s=\text{tr}[A\pl^s A]=\text{tr}[F\pl^{s-2} F]$. When the interactions are switched on, the currents are no longer conserved $\cons{}J_s=g\, \text{tr}[A\pl^{s-1} A^2]+O(g^2)=g\, \text{tr}[F\pl^{s-2} FA]+O(g^2)$. This type of breaking was studied in the case of $\mathcal{N}=4$ SYM in \cite{Anselmi:1998ms,Eden:2004ua,Henn:2005mw,Belitsky:2007jp}.

\paragraph{Quantum, GPZ.} The second mechanism can be called the quantum breaking \cite{Girardello:2002pp,Maldacena:2012sf}. On the CFT side the non-conservation is governed by double- and/or triple-trace operators built out of the HS currents $J$ themselves:
\begin{align}
\cons{} J&=g JJ\text{ possibly plus } g^2JJJ\,.\label{recombinationeq}
\end{align}
This equation is true not only in perturbation theory where the operators renormalize order by order, but at the full quantum level. The simplest consequences of this non-conservation condition we will explore below. It is not clear what should happen in the bulk. In \cite{Girardello:2002pp} it was proposed that the HS fields acquire the mass via loops, which results in a jump in the number of physical degrees of freedom. A delicate UV/IR interplay is required and to see how this is realized on the $AdS$ side is an open problem.

\subsection{Insertions of Equations of Motion}
In this Section we insert the equations of motion into various correlation functions to read off the relations between the CFT data. The main idea is to take advantage of the conformal symmetry when extracting information from the non-conservation of the HS currents. While the HS currents are exactly conserved in the free CFT or at $N=\infty$ in Wilson-Fisher CFT, apart from $N=\infty$ the short multiplets of HS currents recombine with the long multiplets formed by double- and triple-trace operators as to form long multiplets, which is what \eqref{recombinationeq} describes. This phenomenon originates from the recombination of $\phi$ itself, whose structure depends on the quantum equations of motion \cite{Rychkov:2015naa}. We do not consider all possible correlation functions that can be of interest, but only those that talk to the anomalous HS currents.

\subsubsection{Anomalous Two-Point Function of Scalar Field}
The simplest case is to consider the two-point function of the spin field \cite{Rychkov:2015naa}:
\begin{align}
\langle\phi^i \phi^j\rangle=\delta^{ij}\frac{C_{\phi\phi}}{(x^2_{12})^{\Delta_\phi}}\,,
\end{align}
where $\Delta_\phi=d/2-1+\gamma_\phi$. Then we impose the equations of motion abstractly, without having to specify the composite operator that becomes the descendant $\square\phi$:
\begin{align}
\square_1\square_2\langle\phi \phi\rangle&=4 \gamma_\phi  (\gamma_\phi +1) (2 \gamma_\phi +d-2) (2 \gamma_\phi +d) \frac{C_{\phi\phi}}{(x^2_{12})^{\Delta_\phi+2}}\,.
\end{align}
Another way to compute the same is to drag $\square$ inside the correlator and replace it with the composite operator it is supposed to recombine with. This depends on the model. In the large-$N$ Wilson-Fisher $\square \phi^i=g_\star \sigma \phi^i$ and via the Wick theorem we can get to the lowest order:
\begin{align}
\square_1\square_2\langle\phi^i \phi^j\rangle&=g_\star^2 \langle (\sigma \phi^i)(\sigma\phi^j)\rangle=\delta^{ij}g_\star^2 \frac{C_{\sigma\sigma}C_{\phi\phi}}{(x_{12}^2)^{\Delta_\phi+2}}\,,
\end{align}
where there is a critical coupling $g_\star$ and the $\sigma$-field two-point function normalization $C_{\sigma\sigma}$. Therefore, to the lowest order in perturbation theory we find a relation between the three quantities:
\begin{align}
C_{\sigma\sigma} g_\star^2 &= 4d(d-2)\gamma_\phi\label{phisigma}\,.
\end{align}
This relation is indeed true as can be seen from Appendix \ref{app:models}. One more example is the $\phi^3_6$-theory where $\square \phi=g_\star \phi^2$. It is formally the same as \eqref{phisigma} where $d=6$ and $\sigma=\phi$. Analogously, in the $4-\epsilon$ Wilson-Fisher case we have $\square \phi^i=g_\star \phi^i(\phi)^2$ and
\begin{align}
\square_1\square_2\langle\phi^i \phi^j\rangle&=g_\star^2 \langle (\phi^i\phi^2)(\phi^j\phi^2))\rangle=\delta^{ij}2(N+2)g_\star^2  \frac{C_{\phi\phi}^2}{(x_{12}^2)^{3\Delta_\phi}}\,,
\end{align}
which results in
\begin{align}
\gamma_\phi &=\frac{(N+2)g_\star^2 }{256 \pi^4}\label{gammaphiWF}\,.
\end{align}
In practice it is convenient to measure the anomalous dimensions of composite operators in terms of that of $\phi$ rather than less physically interesting quantities like $C_{\sigma\sigma}$ and $g_\star$. Eq. \eqref{phisigma}, \eqref{gammaphiWF} allows us to do so since it is $C_{\sigma\sigma} g_\star^2$ that shows up in the formulas. The insertions of equations of motion was successfully applied recently in \cite{Rychkov:2015naa,Raju:2015fza,Ghosh:2015opa}.

\subsubsection{Anomalous Two-Point Function of HS Currents}

Interactions can break the HS symmetry and, as a result, the HS currents are no longer conserved and develop an anomalous dimension $\gamma=\gamma_s$
\begin{align}
\langle J_s(x_1,\xi_1) J_s(x_2,\xi_2)\rangle&=\frac{C_{J,s}}{\mu^{2\gamma}(x_{12}^2)^{d+s-2+\gamma}}(P_{12})^s \label{anomtwopointJJ}\,,
\end{align}
where $C_{J,s}$ is the spin-$s$ central charge and $\mu$ is the renormalization scale to compensate for the anomalous dimension $\gamma$, which we drop in what follows.

The main idea of the Anselmi's trick \cite{Anselmi:1998ms} is that one can check the conservation of the anomalous HS current inside the two-point function in two different ways. First of all, one can take double-divergence of the two-point function as a whole, i.e. to check the conservation of \eqref{anomtwopointJJ}, the result being proportional to the anomalous dimension. Secondly, one can drag the double-divergence inside the two-point function, i.e. replace $\cons{}J$ with its expression in a given theory. This can be written as
\begin{align}\label{anselmieq}
{\langle \cons{1}J_s(x_1)\cons{2}J_s(x_2)\rangle}&={\cons{1}\cons{2}\langle J_s(x_1) J_s(x_2)\rangle}\,.
\end{align}
The identity above becomes a source of nontrivial information when the two sides can be computed independently. For example, the non-conservation is governed by some small parameter $g_\star$:
\begin{align}
\cons{} J_s= g_\star K_{s}\,.
\end{align}
Then, the left-hand side of \eqref{anselmieq} is proportional to $(g_\star)^2$, while the right-hand side can be explicitly evaluated from \eqref{anomtwopointJJ}. One observes that the anomalous dimension is by $(g_\star)^2$ closer than before --- one needs to evaluate $\langle KK\rangle$ and $\langle JJ\rangle$ to $n$ loops in order to determine the anomalous dimension at loop order $n+1$. In particular, it is enough to work with an almost 'classical' theory in order to find anomalous dimensions to the one-loop order. One more advantage of the method is that it allows to treat by the same means different theories in diverse dimensions that are originally studied by distinct methods. Let us compute the right-hand side of \eqref{anselmieq} by taking the divergence at both the insertions ($h=d/2$):
\begin{align}\label{nonconservationtwopoint}
\cons{1}\cons{2}\langle J_s(x_1) J_s(x_2)\rangle&=\frac{(P_{12})^{s-2}}{(x_{12}^2)^{d+s-1+\gamma}}\Big[ P_{12} a(\gamma)+\tilde{P}_{12} b(\gamma)\Big]\,,\\
a(\gamma)&=2 \gamma  s (h+s-2) \left(2 h^2+2 \gamma  (h+s-2)+3 h s-5 h+s^2-4 s+3\right)\,,\\
b(\gamma)&=-2 \gamma ^2 (s-1) s (h+s-3) (h+s-2)\,,
\end{align}
where there is an additional conformal structure
\begin{align}
\tilde{P}_{12}&=2\frac{(x_{12}\cdot \xi_1)(x_{12}\cdot \xi_2)}{x_{12}^2}\,.
\end{align}
The HS current is a primary operator by definition, but this is not the case for its divergence, which is a typical descendant. The new structure $\tilde{P}_{12}$ appears since the two-point functions of descendants are more complicated. When the anomalous dimension $\gamma$ vanishes the divergence becomes a primary operator that decouples and in this case the HS current is exactly conserved, which explains the factors of $\gamma$ and $\gamma^2$ in front of the usual conformal structure $P_{12}$ and the unusual one $\tilde{P}_{12}$.

Since the space-time dependence of \eqref{nonconservationtwopoint} is fixed, it is convenient to take it at $x_1=x$, $x_2=0$, $\xi_1=\xi_2=\xi$, which results in \cite{Belitsky:2007jp}
\begin{align}
\cons{1}\cons{2}\langle J_s(x_1) J_s(0)\rangle&=\left(\frac{-2(\xi\cdot x)^2}{x^2}\right)^{s-1} \frac{1}{(x^2)^{d+s-1+\gamma}} c(\gamma)\,,\\
c(\gamma)&=2 \gamma  s (h+s-2)(\gamma  (h s+h+(s-2) s-1)+(h+s-1) (2 h+s-3))\,.
\end{align}
Therefore, to the lowest order we find
\begin{align}
c(\gamma) &=\gamma c_1+O(\gamma^2)\,, &&c_1=\frac{1}{2} s (d+s-3) (d+2 s-4) (d+2 s-2)\,.\label{kinfactor}
\end{align}
The two-point function of $K$ has the same space-time dependence and is defined by a number $C_{K,s}$
\begin{align}
\langle K_s(x_1)K_s(x_2)\rangle&=C_{K,s}\left(\frac{-2(\xi\cdot x)^2}{x^2}\right)^{s-1} \frac{1}{(x^2)^{d+s-1+\gamma}}\,.
\end{align}
Finally, the equation \eqref{anselmieq} for the anomalous dimension to the lowest order is
\begin{align}\label{mainformula}
g^2_\star\frac{C_{K,s}}{C_{J,s}}&=c(\gamma)=g^2_\star c_1 \gamma_1+\text{higher orders}\,,
\end{align}
where in the last expression we assume that the anomalous dimension is expanded as $\gamma=g^2_\star\gamma_1+...$. Next, we will derive $K$ for a number of models and compute $\gamma_1$.

\subsection{Non-conservation of HS Currents}
Below, we give the explicit form of the composite operators responsible for the non-conservation of HS currents in some bosonic models that are described in Appendix \ref{app:models} in more detail.

\paragraph{General Formulae.}
The HS currents are no longer conserved in the interacting vector model, the non-conservation having the form of double-trace operators for the bosonic models:
\begin{align}
\cons{}J=g JJ\,.
\end{align}
Explicitly, using the generating function \eqref{HScurrentscalar} we find:
\begin{align}
\cons{} J_s(x,\xi)=(\xi\cdot \pl_1+\xi\cdot \pl_2)^{s-1}\left[f_1(u)\square_1 \phi(x_1)^i\phi^j(x_2)+
f_2(u) \phi(x_1)^i\square_2\phi^j(x_2)\right]\Big|_{x_1=x_2=x}\,,
\end{align}
where $u=(\xi\cdot \pl_1-\xi\cdot \pl_2)/(\xi\cdot \pl_1+\xi\cdot \pl_2)$ and the $\pl_1\cdot \pl_2$-terms vanish because $J$ was designed as to ensure $\cons{} J(x,\xi)=0$ whenever $\square_{1,2}=0$. The expression in the square bracket can be worked out without using the explicit form of the quantum equations of motion. It can be simplified with the help of the recurrence relations that the Gegenbauer polynomials obey with the result:\footnote{These formulae agree for $d=6$ with \cite{Belitsky:2007jp}.}
\begin{align}
K&=f_1(x)+f_2(y)\,,
&&\begin{aligned}
f_1(x)&=+4 \nu  (\nu +1) (x-1) C_{s-2}^{(\nu +2)}(x)+\nu  (2 \nu +1) C_{s-1}^{(\nu +1)}(x)\,,\\
f_2(y)&=-4 \nu  (\nu +1) (y+1) C_{s-2}^{(\nu +2)}(y)-\nu  (2 \nu +1) C_{s-1}^{(\nu +1)}(y)\,,
\end{aligned}\end{align}
and the variables $x$ and $y$ point split the equations of motion of interest. Note that $K(x,y)$ inherits the symmetry of the original currents under the permutation of the arguments, i.e. $K(-y,-x)=(-)^{j} K(x,y)$ and $f_2(-x)$ is $(-)^sf_1(x)$.

\paragraph{Large-$\boldsymbol{N}$.} For the sigma model the source is bilinear:
\begin{align}
\square \phi^i=g_\star  \phi^i\sigma\,.
\end{align}
Here the double-trace operator $K_s$ measuring the non-conservation is:
\begin{align}\notag
K_s=\cons{} J_s(x,\xi)&=g_\star(\xi\cdot \pl_1+\xi\cdot \pl_2+\xi\cdot\pl_3)^{s-1}\, K_s(x,y) \,\phi^i(x_1)\phi^j (x_2)\sigma(x_3) \Big|_{x_i=x}\,,
\end{align}
where
\begin{align}\label{triplesplit}
x&=\frac{\xi\cdot \pl_1-\xi\cdot \pl_2+\xi\cdot\pl_3}{\xi\cdot \pl_1+\xi\cdot \pl_2+\xi\cdot\pl_3}\,, &
y&=\frac{\xi\cdot \pl_1-\xi\cdot \pl_2-\xi\cdot\pl_3}{\xi\cdot \pl_1+\xi\cdot \pl_2+\xi\cdot\pl_3}\,.
\end{align}
The same expression works fine for the $6d$ $\phi^3$-theory with the understanding that $\sigma$ is one of the $\phi$ fields whose two-point function is that of the free scalar field to the lowest order, while for the large-$N$ vector model $G_\sigma^{-1}=-\tfrac12 G_\phi^2$.

The form of the non-conservation operator just obtained is sufficient to do computations, but it is also instructive to derive the decomposition in terms of composite operators $JJ$ built of the single-trace operators: \begin{align}\label{JJsigmamodel}
\cons{} J_s&= g_\star \sum_{a+c<s} f_{a,c}\,\pl^a J_{s-1-a-c}\,\pl^c \sigma\,.
\end{align}
Directly re-expanding the generating function $K$ we find (recall that $\nu=(d-3)/2$)
\begin{align}
f_{a,c}&=-\frac{2 (a+c+1)! (a+c-\nu -s+1) (a+2 c+1-2 (s+\nu ))!}{a! c! (c+1)! (a+c-2 (s+\nu ))!}\,.\label{ClebshGordonN}
\end{align}
Eq. \eqref{JJsigmamodel} is helpful in evaluating the two-point function $\langle KK\rangle$ as it diagonalizes $J$ and $\sigma$. For the case of the spin-four non-conservation operator we find $f_{1,0}/f_{0,1}|_{s=4}=-2/5$ as in \cite{Maldacena:2012sf,Giombi:2011kc}:
\begin{align}
\cons{}J_4\sim J_2\,\pl \sigma-\frac{2}{5}\pl J_2\, \sigma\,.
\end{align}
The main source of the HS symmetry breaking in \cite{Maldacena:2012sf,Giombi:2011kc} was the violation of parity coming from the Chern-Simons coupling. Note that the formulae derived above apply strictly speaking only to the case of non-singlet currents as for the singlet ones it yields the terms of the form $J_0\sigma$ and there is no $\phi^2$ operator, but there is $\phi^i\phi^j$, $i\neq j$.

\paragraph{Wilson-Fisher $\boldsymbol{4-\epsilon}$.} For the Wilson-Fisher model in $4-\epsilon$ the source is trilinear:
\begin{align}
\square \phi^i=g_\star \phi^i \phi^2\,,
\end{align}
which leads to
\begin{align}\label{fourepsnoncon}
K_s=\cons{} J_s(x,\xi)&=g_\star\left(\sum_{k=1}^{4}\xi\cdot \pl_k\right)^{s-1}\, K_s(x,y) \,\phi^i(x_1)\phi^j (x_2)\phi^k(x_3) \phi_k(x_4)\Big|_{x_i=x}\,,
\end{align}
where $K(x,y)$ is the same function as before, but the arguments have to be changed to
\begin{align}\label{quarticsplit}
x&=\frac{\xi\cdot \pl_1-\xi\cdot \pl_2+\xi\cdot\pl_3+\xi\cdot\pl_4}{\xi\cdot \pl_1+\xi\cdot \pl_2+\xi\cdot\pl_3+\xi\cdot\pl_4}\,, &
y&=\frac{\xi\cdot \pl_1-\xi\cdot \pl_2-\xi\cdot\pl_3-\xi\cdot\pl_4}{\xi\cdot \pl_1+\xi\cdot \pl_2+\xi\cdot\pl_3+\xi\cdot\pl_4}\,.
\end{align}
Again one can diagonalize the non-conservation operator and represent it as $JJ$:
\begin{align}
\cons{} J^{ij}_s&=\sum_{a+c+s'=s-1} A_{a,c}^{s'}\, \pl^a J^{ij}_{s'}\, \pl^c J_{0}\,,
\end{align}
where the second factor is always $J_0=\phi^2$ and the coefficients are
\begin{align}\label{ClebshGordonEps}
A_{a,c}^{s}&=\frac{2 (-1)^a (\nu +s) (a+c+1)! (a+c+2 \nu +2 s+1)!}{a! c! (c+1)! (a+2 (\nu +s))!}\,.
\end{align}
$A_{a,c}^{s}$ vanishes whenever $a+c$ is even. For example, the non-conservation operator for the spin-two current
\begin{align}
\cons{}J^{ij}_2&= a_1\pl J_0^{ij}J_0+a_2J^{ij}_0\pl J_0\,, & a_1&=-8 \nu  (\nu +1)\,, &a_2&=4 \nu  (\nu +1) (2 \nu +1)\,,
\end{align}
is non-trivial for the non-singlet sector, but vanishes for the singlet one at $d=4$ since $a_2=-a_1=6$ for $\nu=\tfrac12$, i.e. the traceless stress-tensor is conserved as expected.\footnote{There is, of course, a conserved stress-tensor for any $d$ in the $\phi^4$ theory, but it is not traceless and the traceless one is conserved only in $d=4$.}

\subsection{Anomalous Dimensions}
Below we combine the formulae for the non-conservation of the HS currents derived in the previous Section with the Wick theorem as to get the anomalous dimensions $\gamma_s$ of the HS currents to the one-loop order. The main formula is \eqref{mainformula}, where $C_{J,s}$ was computed in Section \ref{sec:Unbroken}. In addition \eqref{phisigma} and \eqref{gammaphiWF} allows us to express $\gamma_s$ in the units of $\gamma_\phi$. The only missing ingredient is $C_{K,s}$, i.e. the two-point functions $\langle KK\rangle$.

\subsubsection{$\boldsymbol{4-\epsilon}$}

To determine the anomalous dimension of the non-singlet currents we need to compute two contributions --- two distinct Wick contractions:
\begin{align}
\langle K^{12} K^{12}\rangle&= 2N \langle 1234 \rangle+8 \langle1324\rangle\,,
\end{align}
where we indicated the $N$ factors explicitly as well as the permutation of the $\phi$'s in \eqref{fourepsnoncon}. Also, we took only the $K^{12}$ component as a generic non-singlet. The correlator $\langle K_jK_j\rangle$ is expected to have the form
\begin{align}
I_{\sigma}&=C_{K,j} \frac{  [-2(\xi\cdot x)^2]^{j-1}}{(x^2)^{2d+2j-6}}\,,
\end{align}
which should be applied to the $d=4$ case. The prefactor splits as
\begin{align}
C_{K,j}&=\frac{C_{\phi\phi}^4\nu^2 2^{j-1}\Gamma[2d+2j-6]}{\Gamma[d/2-1]^4} \times I\,,
\end{align}
where $C_{\phi\phi}$ is the $\phi$ two-point function normalization \eqref{phitwopoint} and $I$ is the integral over the Schwinger parameters that depends on the permutation. We only need:
\begingroup\allowdisplaybreaks
\begin{align}
I_{1234}&=\frac{(j-1) j (j+1) (j+2)}{2 j+1}\,, &
I_{2314}&=\frac{j (j+1) \left((-)^j( j^2+j-4 )-2\right)}{2 (2 j+1)}\,,\\
I_{1324}&=I_{3214}=(-)^j I_{2314}\,,  &
I_{3412}&= -\frac{2  \left((-)^j+1\right) j (j+1)}{2 j+1}\,.
\end{align}
\endgroup
One has to distinguish between odd and even spins. The relation \eqref{gammaphiWF} between the coupling constant $g_\star^2$ and anomalous dimension of the spin field $\gamma_\phi$ allows us to express everything in the units of $\gamma_\phi$:
\begin{align}
\gamma^{\YoungmB} &= 2\gamma_\phi\left(1-\frac{2(N+6)}{(N+2)s(s+1)}\right)\,,\\
\gamma^{\YoungmAA} &= 2\gamma_\phi\left(1-\frac{2}{s(s+1)}\right)\,,\label{ONcurrent}
\end{align}
where the only difference between the two cases comes from whether the current has even or odd spin. This is the correct formula that dates back to Wilson and Kogut \cite{Wilson:1973jj}, see also \cite{Braun:2013tva} for the more clean decomposition into $O(N)$ irreducible structures and higher orders. Note that \eqref{ONcurrent} vanishes for $s=1$ since it is the global $O(N)$-symmetry current. For the singlet HS currents we find the similar permutations but with different prefactors:
\begin{align}
\langle K^{\bullet} K^{\bullet}\rangle\sim 4N^2 \langle 1234 \rangle+16N \langle2314\rangle+4N^2\langle3412\rangle\,,
\end{align}
which again leads to Wilson and Kogut:
\begin{align}
\gamma &= 2\gamma_\phi\left(1-\frac{6}{s(s+1)}\right)\,.
\end{align}
The anomalous dimension vanishes for the stress-tensor $s=2$, as is expected.

\subsubsection{Large-$N$}
In the case of the large-$N$ expansion first of all we strip off the trivial factors from $\langle KK\rangle$:
\begin{align}
\langle K_jK_j\rangle&=C_{K,j} \frac{  [-2(\xi\cdot x)^2]^{j-1}}{(x^2)^{d+2j-2}}\,.
\end{align}
After simple manipulations we find ($\Delta_\phi=d/2-1$, $\Delta_\sigma=2$)
\begin{align}
C_{K,j}&=\nu^2 \frac{ C_{\phi\phi}^2C_{\sigma\sigma}}{\Gamma[\Delta_\phi]^2\Gamma[\Delta_\sigma]}\Gamma[d+2j+\Delta_\sigma-4]2^{j-d-\Delta_\sigma+2} \times I\,,
\end{align}
where $I$ is the integral over the Schwinger parameters:
\begin{align}
I\Big|_{\Delta_\sigma=2}&=\frac{32 \sqrt{\pi } (d-1) \Gamma \left(\frac{d}{2}-1\right) \Gamma (d+s-1)}{(d+2 s-3) \Gamma \left(\frac{d-1}{2}\right) \Gamma (d+1) \Gamma (s-1)}\,.\label{largenintegral}
\end{align}
Using \eqref{phisigma} we can write the anomalous dimensions as a multiple of $\gamma_\phi$:
\begin{align}\label{nonsinnglets}
\gamma_s&=\gamma_\phi\frac{8 (s-1) (d+s-2)}{(d+2 s-4) (d+2 s-2)}\,,
\end{align}
which is in accordance with Lang and Ruhl \cite{Lang:1990re}. This formula gives $\gamma_s^{\YoungmAA}$ for odd $s$ and $\gamma_s^{\YoungmB}$ for even $s$. The same formula is applicable to the large-$N$ expansion of the recently studied $O(N)$ vector-model in $4<d<6$ \cite{Bekaert:2011cu,Fei:2014yja,Fei:2014xta,Gracey:2015tta}. It vanishes for $s=1$ as before.

\subsubsection{Six Dimensions and Nearby}
Another useful toy models are the six-dimensional $\phi^3$ theories
\begin{align}
S=\int d^dx\,\left(\frac12(\pl\phi)^2+\frac{g}{2}d^{IJK}\phi_I\phi_J\phi_K\right)\,,
\end{align}
for various $d^{IJK}$, see e.g. \cite{deAlcantaraBonfim:1980pe,Braun:2013tva}, of which we consider two particular cases: the scalar QCD \cite{Baulieu:1979mr,Mikhailov:1984cp}
\begin{align}\label{sixdqcd}
S=\int d^dx\,\left(\frac12(\pl\phi)^2+\frac12(\pl\sigma)^2+\frac{g}{2}\phi_i\phi^i\sigma\right)\,,
\end{align}
whose $\beta$-function can be made to vanish up to $g^5$ by adjusting the field content; also we include the recently studied UV completion of the $O(N)$ vector-model in $4<d<6$ \cite{Fei:2014yja,Fei:2014xta,Gracey:2015tta}
\begin{align}\label{uvcompletion}
S=\int d^dx\,\left(\frac12(\pl\phi)^2+\frac12(\pl\sigma)^2+\frac{g_1}{2}\phi_i\phi^i\sigma+\frac{g_2}{6}\sigma^3\right)\,.
\end{align}
For simplicity we restrict ourselves to the non-singlet currents built out of $\phi^i$. In six dimensions we have $C_{\sigma\sigma}=C_{\phi\phi}=\Gamma[d/2-1]/(4\pi^{d/2})$, $\Delta_\sigma=\Delta_\phi=d/2-1=2$. From the computational point of view the only difference between this model and the large-$N$ vector-model is in that the normalization of the $\sigma$-field is that of the free field. Therefore, we can directly use \eqref{largenintegral} at $d=6$. For the anti-symmetric $O(N)$-representation (the same for the symmetric) we find
\begin{align}
\gamma_s&=2\gamma_\phi\frac{(s-1) (s+4)}{(s+1) (s+2)}=12\gamma_\phi\left(\frac16 -\frac{1}{(s+1)(s+2)}\right)\,,
\end{align}
which is in accordance with \cite{Baulieu:1979mr,Mikhailov:1984cp,Belitsky:2007jp,Manashov:2015fha}. The anomalous dimension vanishes at $s=1$ for the current is that of the global $O(N)$-symmetry. The same formula applies to the  $6-\epsilon$ expansion of \eqref{uvcompletion} and to \eqref{sixdqcd} at $6d$.

\subsection{HS Algebra Interpretation}
Below we vaguely argue how the anomalous dimension of the HS currents at one-loop order are related to the HS algebra representation theory. Despite not having anything to do with loop integrals the computations above reveal that the anomalous dimensions are simpler than some of the ingredients used to derive them: the two-point functions of the HS currents almost cancel those of the non-conservation operators yielding simple rational expressions.

Let us recall that the HS currents multiplet $\mathbb{J}=\sum_s J_s$ forms an irreducible representation $\mathbb{J}$ of the HS algebra corresponding to a free field they are built from. Therefore, the non-conservation operator $K$ belongs to its tensor square for the case of the $4-\epsilon$ expansion, where we depart from the free theory,
\begin{align}
K=\cons{} J \in  \mathbb{J}\otimes \mathbb{J}
\end{align}
In fact, the tensor product we need is $\mathbb{J}\otimes J_0$. By definition, the generating function we found for $K$ can be re-expanded in terms of the Clebsh-Gordon coefficients of the HS algebra, which is \eqref{ClebshGordonEps}. For the case of the $6-\epsilon$-expansion we need instead $\mathbb{J}\otimes \phi$.

Another important ingredient is the kinematical factor $c_1$ \eqref{kinfactor} that comes from decoupling the divergence of the HS current inside the two-point function. This tells us that we cannot express everything in terms of the HS algebra associated with the free boson, for which the HS currents are strictly conserved. To get this factor right one needs the HS algebra for generalized free fields near the unitarity bound. Such HS algebra is defined \cite{Alkalaev:2014nsa} as a quotient of the universal enveloping algebra $U(so(d,2))$ of the conformal algebra by certain ideal:
\begin{align}
\text{hs}(\lambda)&= U\left(\YoungpAA\right)/ I_\lambda\,,&& I_\lambda= \YoungpAAAA\oplus (C_2-C_2(\Delta))\,,\label{genff}
\end{align}
where \YoungmAA{} stands for the $so(d,2)$ generators $T_{AB}=-T_{BA}$ and the first part of the ideal is generated by the totally anti-symmetric bilinear $T_{[AB}T_{CD]}$. The first factor of the ideal implies that the field is a scalar one, while the second factor fixes the value of the Casimir operator $C_2$ to be that of the generalized free field of weight-$\Delta$, i.e. $C_2(\Delta)=\Delta(\Delta-d)$. We are interested in $\Delta=(d-2)/2+\epsilon$. At $\epsilon=0$ one finds an additional ideal in $\text{hs}(\lambda)$ and the quotient is the Eastwood-Vasiliev HS algebra \cite{Eastwood:2002su,Vasiliev:2003ev} that corresponds to the free boson. In all the cases the correlation functions of $J$'s can be computed as HS invariants \cite{Colombo:2012jx,Didenko:2012tv}, the same for $J\otimes J$ etc. as we argue in Appendix \ref{app:Duals}. In addition, the algebra \eqref{genff} determines the actual coefficients that multiply the independent singular vectors that enter $K$.

The interpretation of the large-$N$ limit is less straightforward. The auxiliary weight-two field is not a free one, except for $d=3$ where one can think of $\sigma$ as $\bar{\psi}\psi$ for the purpose of deriving the Clebsh-Gordon coefficients. In generic dimension one has to take the representation that is dual to $\mathbb{J}$, which will give $\tilde{J}_0$ of weight-two.

It would be important to reformulate the entire procedure in terms of the higher-spin symmetries with the goal to reveal the right algebraic structure governing the HS currents subsector of Wilson-Fisher CFT's.

\section{AdS}\label{sec:ads}

In this section we discuss various implications for the AdS dual HS theories
that can be extracted from the CFT results obtained above. Firstly, we reconstruct the
part of the dual HS theory that produces the required three-point functions $\langle J_s OO\rangle$ or better to say the invariants $I_{s00}$ \eqref{HSinvariant}. Secondly, we discuss the expected one-loop corrections to the $AdS$-masses of HS fields. In Appendix \ref{app:Duals} we show how to enlarge HS theories with the duals of the multi-trace operators.

\subsection{Cubic Couplings with Two Scalar Legs}

The knowledge of the correlation functions of some CFT can be used to reconstruct the dual AdS theory, see e.g. \cite{ElShowk:2011ag} for the recent study. The reconstruction approach should work at least at the level of trees since the kinematical volume is the same on the CFT and AdS sides. For example, the number of independent cubic vertices in $AdS_{d+1}$ \cite{Metsaev:2005ar} is equal to the number of independent three-point structures in CFT$_{d}$ \cite{Costa:2011mg}. The four-point functions of the scalar operators are fixed up to a function of two conformally-invariant cross-ratios, which is in accordance with the quartic vertices in $AdS$ having a doubly infinite expansion in derivatives \cite{Bekaert:2015tva}. The explicit check up to the quartic order on the example of the free boson vs. a subsector of the $4d$ HS theory was done in \cite{Bekaert:2015tva}, where the crucial observation was that the $AdS$ exchange diagrams reproduce the contribution of the single-trace operators to the four-point function, with the remnant corresponding to the double-trace operators accounted for by the quartic contact vertices.

One can start with a given free CFT, compute the three-point invariant $I_{s00}$ \eqref{HSinvariant} and manufacture the interaction vertex in $AdS$ as to reproduce the correlation functions of that CFT. In the case of the $\langle J_s O_\Delta O_\Delta\rangle$ correlator the relevant part of the bulk action is
\begin{align}\label{adsaction}
S&=\sum_s\int \frac12 \big[\Fron_{\mm(s)}(\square- m^2_s)\Fron^{\mm(s)}+...\big] +\sum_s g_s \int \Fron_{\mm(s)} J^{\mm(s)}=S_0+S_1\,,\\
&m^2_{s>0}=\Lambda[(d+s-2)(s-2)-s]\,, \qquad  \qquad\qquad m^2_0=\Delta(\Delta-d)\,,\label{FronsdalMass}
\end{align}
where we truncated the Fronsdal action to its Klein-Gordon part since it is well-known. The first part makes the free action $S_0$, while the second one makes the interactions $S_1$. The interaction is due to the currents $J^{\mm(s)}$ that the HS fields couple to. The currents are built of the scalar field $\Fron_0$ that is a part of the HS multiplet and is included in \eqref{adsaction} as $s=0$. The form of the current can be changed by adding improvements
\begin{align}\label{bulkHScurrent}
J_{\mm(s)}&=\Fron_0(x)\left(\overleftrightarrow{\nabla}_\mm\right)^s\Fron_0(x)+O(g_{\mm\mm},\Lambda)\,,
\end{align}
where we drop the terms with the $AdS_{d+1}$ metric $g_{\mm\mm}$ and cosmological constant $\Lambda$ since the boundary-to-bulk propagator for the HS fields \cite{Mikhailov:2002bp} is naturally traceless. For the purpose of computing the simplest Witten diagram \scalebox{1.5}{$\Yup$} we can pass to the simplified interaction:
\begin{align}
\int \Fron_{\mm(s)} J^{\mm(s)}\sim \int \Fron_{\mm(s)} \nabla^{\mm(s)}\Fron_0 \Fron_0\,.
\end{align}
Therefore, the part of the cubic action that is responsible for $\langle J_s O_\Delta O_\Delta\rangle$ is:
\begin{align}\label{adsactionB}
S&=\sum_s \int \frac12 \big[\Fron_{\mm(s)}(\square- m^2_s)\Fron^{\mm(s)}+...\big] +\sum_s g_s \int \Fron_{\mm(s)} \nabla^{\mm(s)}\Fron_0 \Fron_0\,.
\end{align}

The only information needed from the $AdS$ side is the coefficient of the standard three-point function $\JOOst{s}{\Delta}$ \eqref{standardthreepoint} that is produced from the standard bulk vertex, which was done in \cite{Costa:2014kfa} with the result
\begin{align}
\int \Fron_{\mm(s)} &\nabla^{\mm(s)}\Fron_0 \Fron_0= \tilde{b}_{s00}\times \JOOst{s}{\Delta}\,,\\
\tilde{b}_{s00}&=\frac{2^{-5+2 s} \pi ^{-d/2} (-3+d+2 s)\Gamma\left[-1+\frac{d}{2}+s\right]^3\Gamma[-3+d+s]\Gamma[-1+s+\Delta ]^2}{\Gamma[-2+d+2 s]^2\Gamma[\Delta ]^2}\,,
\end{align}
which corresponds to the unit normalization of the two-point functions \cite{Costa:2014kfa}. Therefore, $g_s \tilde{b}_{s00}$ should exactly match the CFT invariant $I_{s00}$ \eqref{HSinvariant}. For the cases of the free boson and fermion we find:
\begin{align}
\text{boson}&: && g^B_s=\frac{1}{\sqrt{N}} \frac{\pi ^{\frac{d-3}{4}} 2^{\frac{1}{2} (3 d+s-1)} \Gamma \left(\frac{d-1}{2}\right) }{\Gamma (d+s-3)  }\sqrt{\frac{\Gamma \left(\frac{d-1}{2}+s\right)}{\Gamma (s+1)}}\,,\\
\text{fermion}&: && g^F_s=g^B_s \frac{s}{(d+s-3)}\,.
\end{align}
Because of the relation between $g^B$ and $g^F$ displayed in the last line the duals of the free boson and free fermion are different for $d\neq 3$ and should be one and the same HS theory for $d=3$. It is quite remarkable that the difference between the correlation functions in the two CFT's is compensated by the difference of the same bulk integral with different boundary conditions. The case of the free boson was considered in \cite{Bekaert:2015tva}. At $d=3$ the coupling has an especially simple form:
\begin{align}
g^B_s\Big|_{d=3}=g^F_s\Big|_{d=3}&=\frac{2^{\frac{s+8}{2}}}{\sqrt{N} \Gamma (s)}\,.
\end{align}
There are at least two alternative ways to achieve the same result. Firstly, one can carefully fix the normalization of the two-point functions from the quadratic part of the bulk action, i.e. from the Fronsdal action, as was computed by Metsaev in \cite{Metsaev:2009ym}. Secondly, the Ward identities fix the coupling constants for the CFT three-point functions. Likewise, we can use them to fix the part of the cubic action by studying the bulk counter-part of the Ward identities, as in \cite{Freedman:1998tz} for $s=1$. Moreover, using the fact that the divergence of the HS field boundary-to-bulk propagator with respect to the boundary data is a pure gauge transformation in the bulk \cite{Mikhailov:2002bp}, one can see that the CFT HS Ward identities lead to the AdS HS Ward identities.

At this point it is sensible to ask what is the first test of the tree-level HS AdS/CFT that does not immediately follow from the bulk/boundary Ward identities. Indeed, the Fronsdal gauge transformations $\delta \Phi_s=\nabla \xi_{s-1}$ receive corrections as to make the interaction part $S_1$ of the action gauge invariant. For example, \eqref{adsaction} is gauge invariant provided the scalar field transformations, which are trivial at the free level $\delta_0\Phi_0=0$, get corrected as $\delta_1 \Phi_0=\sum g_s\nabla...\nabla\xi_s\Phi_0$. In general, the condition for the action to be gauge-invariant to the cubic order is
\begin{align}\label{NoetherWard}
\delta_1 S_0+\delta_0 S_1+O(g^2)&=0 && \Longleftrightarrow && \cons{} \langle JJJ\rangle=\langle JJ\rangle\,.
\end{align}
The left-hand side can be seen to be related to the CFT Ward identities on the right. When the Ward identities are applied to the holographic three-point functions $\cons{} \langle JJJ\rangle=\cons{} S_1$, the property of the boundary-to-bulk propagators \cite{Mikhailov:2002bp} turns $\cons{}S_1$ into $\delta_0 S_1$. If $\delta_0 S_1\neq0$ the correction to the gauge transformations $\delta_1$ is needed to ensure the gauge invariance of the action. Therefore, if $\delta_0 S_1\neq0$ there is a relation between $S_1$ and $S_0$. This is how the gauge invariance in the bulk is related to the CFT Ward identities. 

However, if $\delta_0 S_1=0$, which is the case for the abelian vertices of the type Weyl tensor cubed, no relation between $S_1$ and $S_0$ follows. The number of abelian vertices should be in accordance with the number of the CFT structures that are well-defined as distributions and obey the trivial Ward identities, \cite{Erdmenger:1996yc}. Therefore, the abelian part of $S_1$ cannot be determined this way and makes a nontrivial prediction. The HS Ward identities (\ref{NoetherWard}.right) capture the full structure of the HS algebra, which is in accordance with the non-abelian vertices being fixed in terms of the HS algebra structure constants as in  \cite{Fradkin:1986qy,Vasiliev:2001wa,Alkalaev:2002rq,Vasilev:2011xf,Boulanger:2012dx,Skvortsov:2015lja,Kessel:2015kna,Charlotte}. 

It seems that one can fix the complete cubic action of the $4d$ HS theory by employing the old result by Metsaev \cite{Metsaev:1991nb, Metsaev:1991mt}. Namely, the most general ansatz for the $4d$ cubic vertices \cite{Bengtsson:1986kh} in flat space was taken in \cite{Metsaev:1991nb, Metsaev:1991mt} and the closure of the Poincare algebra at the quartic order was studied. It was found that all the cubic couplings are fixed in terms of a single coupling constant with the result:
\begin{align}
g_{s_1,s_2,s_3}&\sim\frac{1}{\Gamma[s_1+s_2+s_3]}\,,\label{cubiccoupling}
\end{align}
where we display the important part of the spin dependence. Remarkably, $g_{s_1,s_2,s_3}$ agrees with $g_{s}$ obtained above for $s_2=s_3=0$. The rationale for why the same formula is expected to work in $AdS_4$ is as follows. The interactions of HS fields contain higher derivatives which are accompanied by negative powers of the cosmological constant --- the crucial part of the Fradkin-Vasiliev mechanism \cite{Fradkin:1987ks, Fradkin:1986qy}, which is at the heart of the common belief that no sensible flat limit exists for HS theories. However, the flat limit exists for cubic vertices \cite{Bekaert:2010hw} --- in such a limit only the highest derivative terms survive and go over into the vertices classified by Metsaev in flat space \cite{Metsaev:1996pd,Metsaev:2005ar}. The flat limit of the quartic vertex should reveal some non-localities and indeed those are present in \cite{Metsaev:1991nb, Metsaev:1991mt}. Therefore, not only the cubic formula \eqref{cubiccoupling} should work but there can be a meaningful flat limit of the HS theory with some mild non-localities present.

\subsection{AdS Masses at One Loop}

Despite some mild pathologies present in $4-\epsilon$ expansion of the Wilson-Fisher (WF) CFT whenever the space dimension is non-integer \cite{Hogervorst:2015akt}, the physical observables are well-defined. The very existence of the $4-\epsilon$ approach suggests that the dual HS theory might be defined in $AdS_{5-\epsilon}$. In principle, there exist $d$-dimensional Vasiliev equations \cite{Vasiliev:2003ev} at any integer $d$. They are difficult to define at fractional space-time dimension due to the HS algebra that is built in. One can expect that whenever some observable can be scalarized it then can easily be extended to any $d$. The duality between $WF_{4-\epsilon}$ and the HS theory in $AdS_{5-\epsilon}$ is complementary to the Klebanov-Polyakov conjecture that deals with the large-$N$ expansion.

In case the bulk counterpart of the quantum HS symmetry breaking is in accordance with \cite{Girardello:2002pp} the one-loop anomalous dimensions re-derived above with the help of the broken HS symmetry can be used to predict the corrections to the $AdS_{5}$ masses of HS fields:
\begin{align}
 \delta m^2_s=-2(s-2)\epsilon +2\epsilon^2  \gamma_\phi\left(1-\frac{6}{s(s+1)}\right)\,,
\end{align}
where the first term is just the $\epsilon$-expansion of the Fronsdal field's mass \eqref{FronsdalMass}. Similar prediction within the $1/N$-duality was given by Ruhl in \cite{Ruhl:2004cf} for the $AdS_4/CFT_3$ duality:
\begin{align}
\delta m^2_s&=4\gamma_\phi(s-2)\,,& \gamma_s\Big|_{d=3}=4\gamma_\phi\frac{(s-2)}{(2s-1)}\,, && \gamma_\phi\Big|_{d=3}=\frac{4}{3 \pi ^2}\,.
\end{align}
Borrowing the result by Ruhl and Lang \cite{Lang:1990re,Lang:1992zw} on anomalous dimensions of the HS currents in any dimension $d$, which can be simplified to
\begin{align}
\gamma_s&=\frac{8\gamma_{\phi }}{(d+2s-4)(d+2s-2)}\left((d+s-2)(s-1)-\frac{\Gamma [d+1]\Gamma [s+1]}{2(d-1)\Gamma[d+s-3]}\right)\,,
\end{align}
one expects to find at one-loop for the mass shift of $AdS_{d+1}$ HS fields:
\begin{align}
\delta m^2_s&= (d+2s-4) \gamma_s\,.
\end{align}
Note that there are two types of contributions to the anomalous dimensions: one is exactly the same as for the non-singlet currents \eqref{nonsinnglets} and another one comes in the Feynman diagrams language from the additional diagrams where the indices can form closed loops, which are absent for the non-singlet case. In other words \cite{Alday:2015eya} first term comes from the $\sigma$-exchange while the second receives contributions from the sea of the HS currents. It would be interesting to see how such contributions can be manufactured in $AdS$.

\section{Conclusions}

Our main conclusions are as follows. It is very likely that the presence of an at least one HS conserved tensor in a $d>2$ CFT implies that the CFT is a free one in disguise. Therefore, what makes the HS symmetry interesting is its breaking \cite{Anselmi:1998ms,Maldacena:2012sf}.

There are at least two different ways to break HS symmetries: classical and quantum, the former is realized in Yang-Mills type theories and the latter in Wilson-Fisher CFT's. The bulk realization of the classical breaking is via the usual BEH mechanism and requires HS theories to be extended with an appropriate 'matter content'. On contrary, the quantum breaking does not seem to require any new fields to be coupled, but it has not been yet observed in detail. Knowing the one-loop anomalous dimensions of the HS currents allows to make predictions for the corrections to the AdS-masses of HS fields. In particular it may be instructive to extend the duality to fractional dimensions where the $4-\epsilon$ Wilson-Fisher CFT is dual to HS theory in $AdS_{5-\epsilon}$.

The quantum breaking is realized in the CFT's with less operators at disposal. In both the cases, there is an exact quantum equation of motion that describes the recombination of HS currents with a non-conservation operator. The difference is in the structure of such an operator. We have assumed that the non-conservation equation is $\pl\cdot J=gJJ$ and have used this equation to extract anomalous dimensions of the HS currents to the lowest nontrivial order.

There is a number of obvious extensions of the present work. First of all, one can extend the treatment of the large-$N$ to the singlet sector as well. One can study the fermionic vector-models and more complicated models of Yukawa type. Also, one can allow for parity-violating non-conservation operators as in \cite{Giombi:2011kc,Maldacena:2012sf}, which occurs in the Chern-Simons matter theories.

It would be important to understand what is the algebraic structure behind the quantum HS symmetry breaking, which should allow to avoid any Feynman diagrams computation at higher orders.

Given a CFT one can attempt to manufacture the AdS vertices as to reproduce the given correlation functions. The success story in the case of the free $3d$ scalar vs. a subsector of the $4d$ HS theory up to some of the quartic vertices is in \cite{Bekaert:2014cea,Bekaert:2015tva}. The reconstruction approach raises at least two questions. First, does it work for any CFT, even such as free CFT's? Second, more specific, is the dual of the free boson the same HS theory that solves the Noether procedure for HS fields? The independent check from the bulk side of the reconstruction approach would be important. It is possible to fix certain parts of the HS theory action directly on the AdS side \cite{Fradkin:1986qy,Vasiliev:2001wa,Alkalaev:2002rq,Vasilev:2011xf,Boulanger:2012dx,Skvortsov:2015lja,Kessel:2015kna}. In all known cases the coupling constants are related to the HS algebra in a simple way. Moreover, the bulk/boundary Ward identities arguments show that such results are not independent from the CFT. The nontrivial test at the cubic level should come from the abelian part of the vertices.

\section*{Acknowledgments}
\label{sec:Aknowledgements}
I would like to thank Alexander Manashov and John Gracey for correspondence. I am indebted to Alexander Manashov, Ruslan Metsaev, Dmitry Ponomarev, Charlotte Sleight, Arkady Tseytlin and Sasha Zhiboedov for the very useful discussions and comments and to Dmitry Ponomarev for proofreading. I also would like to thank the organizers of the 3rd Higher-Spins and Holography conference, Moscow, Russia and International Workshop on Higher Spin Gauge Theories, Singapore, where some of these results were reported. I am also grateful to the organizers of the School and Workshop on Higher Spins, Strings and Dualities at Quintay, December 7-18, 2015, Chile for the warm hospitality during the final stage of this work, which was  supported by Conicyt grant DPI 20140115. The research was supported in parts by the RFBR Grant No 14-02-01172. This work was supported by the DFG Transregional Collaborative Research Centre TRR 33 and the DFG cluster of excellence ”Origin and Structure of the Universe”.

\begin{appendix}
\renewcommand{\thesection}{\Alph{section}}
\renewcommand{\theequation}{\Alph{section}.\arabic{equation}}
\setcounter{equation}{0}\setcounter{section}{0}

\section{Two- and Three-point functions, Thomas derivative}
\label{app:Thomas}
\setcounter{equation}{0}
As is well-known, see e.g. \cite{Fradkin:1996is}, the two- and three-point functions are fixed by conformal symmetry up to few numbers. For the $\langle 00\rangle$, $\langle ss\rangle$, $\langle 000\rangle$ and $\langle s00\rangle$ cases there is a unique conformally-invariant structure and the ambiguity is in overall factor only, which we omit:
\begin{align}
\langle O_{\Delta}(x_1) O_{\Delta}(x_2)\rangle&=\frac{1}{(x_{12}^2)^{\Delta}}\,,\\
\langle O^{a(s)}_{\Delta}(x_1) O^{b(s)}_{\Delta}(x_2)\rangle&=\frac{1}{(x_{12}^2)^\Delta} \big(P^{ab}...P^{ab}-\text{traces}\big)\,,\\
\langle O_{\Delta_1}(x_1) O_{\Delta_2}(x_2)O_{
\Delta_3}(x_3)\rangle&=\frac{1}{(x_{12}^2)^{\frac{\Delta_1+\Delta_2-\Delta_3}{2}}(x_{23}^2)^{\frac{\Delta_2+\Delta_3-\Delta_1}{2}}(x_{13}^2)^{\frac{\Delta_1+\Delta_3-\Delta_2}{2}}}\,,\\
\langle O^{a(s)}_{\Delta_1}(x_1) O_{\Delta_2}(x_2)O_{
\Delta_3}(x_3)\rangle&=\big(Q^a...Q^a-\text{traces}\big)\langle O_{\Delta_1}(x_1) O_{\Delta_2}(x_2)O_{
\Delta_3}(x_3)\rangle\left(\frac{x_{12}^2x_{13}^2}{x_{23}^2}\right)^{\frac{s}{2}}\,, \\
P^{ab}&=I^{ab}(x_{12})\,, \qquad\qquad\qquad I_{ab}(x)=\delta_{ab}-\frac{2x_ax_b}{x^2}\,,\\
Q^a&=\left(\frac{x^a_{21}}{x_{21}^2}-\frac{x^a_{31}}{x_{31}^2}\right)\,.
\end{align}
Throughout the text we use the structures above as the standard ones. Since the primary operators have to be traceless tensors we need either to impose the trace projector explicitly or consistently work on the space of all tensors modulo pure traces. The second option is more appealing and can be implemented by contracting all tensor indices with a light-like polarization vectors $\xi^a$, $\xi^a \xi_a=0$, which takes away the traces:
\begin{align}
P&= P^{ab}\xi^1_a \xi^2_b\,, & Q&=Q^a\xi^1_a\,,
\end{align}
and in the most of the cases we can employ a single polarization vector, $\xi_{1,2}=\xi$.

The price to pay is that in order to release an index one needs to use the Thomas derivative, which is an operator that preserves the equivalence relation $f(\xi)\sim f(\xi)+\xi^2 g(\xi)$, i.e. $\tomd_a(\xi^2 f(\xi))=O(\xi^2)$. It was obtained in \cite{Craigie:1983fb} and by Thomas \cite{Thomas}, see also \cite{Costa:2011mg}:
\begin{align}
\tomd_a=\left(\frac{d}2-1+\xi^m\pl_m^\xi\right)\pl^{\xi}_a-\frac12 \xi_a \Delta^\xi\,.
\end{align}
In order to check the conservation of a HS current we need to combine the usual derivative with the Thomas operator:
\begin{align}
\cons{}O(x;\xi)&= \frac{\pl}{\pl x_a} \tomd_a O(x;\xi)\,.
\end{align}

\section{Models}\label{app:models}
\setcounter{equation}{0}
Below we briefly summarize the definitions and the lowest order results for the models that we consider in the main text.

\subsection{Wilson-Fisher $4-\epsilon$}
The famous Wilson-Fisher model \cite{Wilson:1971dc} is defined by the following action:
\begin{align}\label{WFaction}
S&=\int d^dx\, \left[(\pl\phi)^2 +\frac{g\mu^{\epsilon}}{4}(\phi^2)^2\right]\,.
\end{align}
The one-loop results \cite{Wilson:1973jj} for the $\beta$-function and anomalous dimensions of the operators $\phi$ and $\phi^2$ are:
\begin{align}
\beta&=-\epsilon g+(N+8) \frac{g^2}{8\pi^2}\,, && g_*=\frac{8\pi^2}{N+8}\epsilon\,,\\
\gamma_\phi&=\frac{N+2}{4(N+8)^2}\epsilon^2\,, && \Delta_\phi=\frac{d}2-1+\gamma_\phi\,,\\
\gamma_{\phi^2}&=\frac{N+2}{N+8}\epsilon\,, && \Delta_\phi=d-2+\gamma_{\phi^2}\,,
\end{align}
and one has to replace $d$ with $d-\epsilon$. The anomalous dimensions of the HS currents have been computed in \cite{Wilson:1973jj,Braun:2013tva} and we quote them in the main text.

\subsection{Large-$N$ Vector-Model}
The large-$N$ expansion is obtained by introducing the Hubbard-Stratanovich field into \eqref{WFaction}:
\begin{align}
S=\int d^dx\,\left(\frac12(\pl\phi)^2+\frac{1}{2}\sigma\phi^2- \frac{3N}{2g}\sigma^2\right)\,.
\end{align}
The two-point function of the spin field $\phi^i$ is that of the free field, while the two-point function of the auxiliary field $\sigma$ to the leading order is
\begin{align}
\langle \sigma(x_1) \sigma(x_2)\rangle&=\frac{C_{\sigma\sigma}}{(x_{12}^2)^2}\,, && C_{\sigma\sigma}=\frac{2^{d+2} \sin \left(\frac{\pi  d}{2}\right) \Gamma \left(\frac{d-1}{2}\right)}{\pi ^{3/2} \Gamma \left(\frac{d}{2}-2\right)}\,.
\end{align}
The leading $1/N$ corrections to the dimension of $\phi$ and $\sigma$ \cite{Abe:1972, Vasiliev:1981yc,Vasiliev:1982dc,Lang:1990re,Lang:1992zw,Petkou:1994ad} are
\begin{align}
\Delta_{\phi}&=\frac{d}{2}-1+\frac{1}{N}\gamma_\phi+O\left(\frac{1}{N^2}\right)\,, &
\gamma_\phi&=\frac{2 \sin \left(\frac{\pi  d}{2}\right) \Gamma (d-2)}{\pi  \Gamma \left(\frac{d}{2}+1\right) \Gamma \left(\frac{d}{2}-2\right)}\,,\\
\Delta_{\sigma}&=2+\frac{1}{N}\gamma_\sigma+O\left(\frac{1}{N^2}\right)\,, &
\gamma_{\sigma}&=\frac{4(d-2)(d-2)}{(d-4)} \gamma_\phi\,.
\end{align}
The anomalous dimensions for the HS currents are known to the leading order only for the singlet sector \cite{Lang:1990re,Lang:1992zw} and to $1/N^2$ for the non-singlets \cite{Derkachov:1997ch}. In the main text we use $\square \phi^i=g_\star\sigma \phi^i$, where $g_\star=1$.

\section{Duals of the Multi-Trace Operators}\label{app:Duals}
\setcounter{equation}{0}

The Vasiliev HS theories consists of the gauge fields plus a number of scalar/fermions that belong to the same HS algebra multiplet as the graviton and genuine HS gauge fields. There are several reasons to discuss how to extend these HS theories with matter-like fields: making contact with string theory, whose spectrum is by far larger than the spectrum of any of the HS theories --- constant vs. exponential growth of the number of states; understanding the classical breaking of HS symmetries.

It turns out that one can easily extend the Vasiliev HS theories with the fields that are dual to multi-trace operators. This works at least to the lowest order. HS theories can always be thought of as the duals of some free CFT's. In the latter case there is a free field $\phi$ that corresponds to a representation, say $S$, of the conformal algebra. Then, the HS algebra is the algebra of all endomorphisms of $S$, i.e. $S\otimes S^*$. The single-trace operators correspond to $S\otimes S$. As was illustrated above $S\otimes S$ contains an infinite number of HS currents. In the $AdS_4/CFT_3$ case the HS algebra is the algebra of functions in two pairs of operators obeying canonical commutation relations \cite{Vasiliev:1986qx}
\begin{align}
[Y^A,Y^B]&=2iC^{AB}\,, && A,B,...=1,...,4\,.
\end{align}
Formally, $S\otimes S$ and $S\otimes S^*$ are isomorphic as vector spaces and one can embed the duals of the single-trace operators into the HS algebra, which leads to the twisted-adjoint action \cite{Vasiliev:1988sa}
\begin{align}
d C&=\omega\star C-C\star\pi(\omega)\,,
\end{align}
where $\omega$ is a connection of the HS algebra and $\pi$ is induced by the auto-morphism that flips the sign of the AdS-translation (in the CFT it exchanges translations and inversions).

Multi-trace operators are given by higher products $(S\otimes S)^k$ and it is possible to embed them into $(S\otimes S^*)^k$. To add the duals of multi-trace operators one just needs to take several copies of the same oscillators
\begin{align}
[Y^A_m,Y^B_n]&=2iC^{AB}\delta_{mn} && n,m=1,...,k\,.
\end{align}
Multiple copies of the HS algebra oscillators naturally show up when studying the conformal HS fields in generalized space-times as in \cite{Gelfond:2003vh, Gelfond:2010pm}. The equations for the duals of the multi-trace operators are
\begin{align}
d C_k&=\omega\star C_k-C_k\star\pi(\omega)\,,
\end{align}
where $\omega$ acts on $C_k$ as on the tensor product. Such equation describes the coupling between gauge HS fields and the duals of multi-trace operators to the lowest order. Higher-trace operators are dual to massive fields as the higher tensor products decompose into long representations \cite{Vasiliev:2004cm}.

The lowest component in the order-$k$ multi-trace tower is the dual of $(\phi^2)^k$-type operators, which has to have the conformal weight $\Delta_k=k(d-2)$, i.e. $k$ in $d=3$. This can be easily seen by repeating the derivation of the boundary-to-bulk propagators in \cite{Didenko:2012tv}. In particular, the correlation functions in the free theory can be reproduced as HS invariants, which immediately follows from \cite{Didenko:2012tv}. Such invariants are needed to reproduce the correlation functions of double-trace operators $JJ$ from the HS algebra perspective.

\end{appendix}

\ifpdf%
{\begingroup%
\linespread{1}\selectfont%
\setlength{\emergencystretch}{8em}%
\printbibliography%
\endgroup%
}%
\else%
\bibliographystyle{unsrt}%
\bibliography{megabib}%
\fi%
\end{document}